\documentclass{article}
\usepackage{times,mathptmx}
\usepackage[T1]{fontenc}
\usepackage{textcomp}
\usepackage[agsmcite]{harvard}
\usepackage[dvips]{graphicx}
\usepackage{subfigure,fullpage}

\newcommand{\text}[1]{\mbox{#1}}
\def\func#1{\mathop{\rm #1}\nolimits}%
\usepackage{amsmath}

\newcommand{\QED}{~~\rule[-1pt]{6pt}{6pt}}
\newcommand{\reals}{{\mbox{\bf R}}}
\newcommand{\symm}{{\mbox{\bf S}}}  

\newtheorem{theorem}{Theorem}

\newtheorem{lemma}[theorem]{Lemma}
\newtheorem{notation}[theorem]{Notation}

\newtheorem{proposition}[theorem]{Proposition}

\newenvironment{proof}{\textbf{Proof.}}{\QED\bigskip}
\renewcommand{\cite}{\citeasnoun}

\begin{document}

\title{Interest Rate Model Calibration Using Semidefinite Programming\thanks{%
This work has been developed under the direction of Nicole El
Karoui and I am extremely grateful to her. Every error is of
course mine. I also benefited greatly in my work from discussions
with Marco Avellaneda, Guillaume Amblard, Vlad Bally, Stephen
Boyd, J\'{e}r\^{o}me Busca, Rama Cont, Darrell Duffie, Laurent El
Ghaoui,\ Eric Fourni\'{e}, Haitham Hindi, Jean-Michel Lasry,
Claude Lemar\'{e}chal, Pierre-Louis Lions, Marek Musiela, Ezra
Nahum, Pablo Parillo, Antoon Pelsser, Ioanna Popescu, Yann
Samuelides, Olivier Scaillet, Robert Womersley, seminar
participants at the GdR FIQAM at the Ecole Polytechnique, the May
2002 Workshop on Interest Rate Models organized by Fronti\`{e}res
en Finance in Paris, the AFFI 2002 conference in Strasbourg and
the Summer School on Modern Convex Optimization at the C.O.R.E. in
U.C.L. Finally, I am very grateful to J\'{e}r\^{o}me Lebuchoux,
Cyril Godart and everybody inside FIRST and the S.P.G. at Paribas
Capital Markets in London, their advice and assistance has been
key in the development of this work. }}
\author{A. d'Aspremont\thanks{Research done while at the CMAPX, Ecole Polytechnique, Palaiseau, France. Email:
alexandre.daspremont@polytechnique.org. Currently: MS\&E dept. and
ISL, Stanford University, Stanford CA, USA. Mailing address: 4076
17th \#B, San Francisco, CA 94114, USA.}} \maketitle

\begin{abstract}
We show that, for the purpose of pricing Swaptions, the Swap rate and the
corresponding Forward rates can be considered lognormal under a single
martingale measure. Swaptions can then be priced as options on a basket of
lognormal assets and an approximation formula is derived for such options.
This formula is centered around a Black-Scholes price with an appropriate
volatility, plus a correction term that can be interpreted as the expected
tracking error. The calibration problem can then be solved very efficiently
using semidefinite programming.

\end{abstract}

\section{Introduction}

In the original \cite{Blac73} model, there is a one-to-one correspondence
between the price of an option and the volatility of the underlying asset.
In fact, options are most often directly quoted in terms of their \cite
{Blac73} implied volatility. In the case of options on multiple assets such
as basket options, that one-to-one correspondence between market prices and
covariance is lost. The market quotes basket options in terms of their \cite
{Blac73} volatility but has no direct way of describing the link between
this volatility and that of the individual assets in the basket. Today, this
is not yet critically important in equity markets where most of the trading
in basket options is concentrated among a few index options, we will see
however that it is crucial in interest rate derivative markets where most of
the volatility information is contained in a rather diverse set of basket
options.

Indeed, a large part of the liquidity in interest rate option markets is
concentrated in European Caps and Swaptions and, as always, market operators
are faced with a modelling dilemma: on one hand, the arbitrage-free price
derived from a dynamic hedging strategy \`{a} la \cite{Blac73} and \cite
{Mert73} has become a central reference in the pricing and risk-management
of financial derivatives, on the other hand however, every market operator
knows that the data they calibrate on is not arbitrage free because of
market imperfections. Beyond these discrepancies in the data, daily model
recalibration and the non-convexity of most current calibration methods only
add further instability to the derivative pricing, hedging and
risk-management process by exposing these computations to purely numerical
noise. One of the crucial filters standing between those two sets of prices
(market data and\ computed derivative prices) is the model calibration
algorithm.

Recent developments in interest rates modelling have led to a form of
technological asymmetry on this topic. Theoretically, models such as the
Libor market model of interest rates (see \cite{BGM97}, \cite{Milt95}) or
the affine Gaussian models (see \cite{elka92} or \cite{Duff96b})\ allow a
very rich modelling and pricing of the basic interest rate options (Caps and
Swaptions) at-the-money. However, due to the inefficiency and instability of
the calibration procedure, only a small part of the market covariance
information that could be accounted for in the model is actually exploited.
To be precise, the most common calibration techniques (see for example \cite
{Long00}) perform a completely implicit fit on the Caplet variances while
only a partial fit is made on the correlation information available in
Swaptions. Because of these limitations, a statistical estimate must be
substituted to the market information on the Forward Libors correlation
matrix as the numerical complexity and instability of the calibration
process makes it impossible to calibrate a full market covariance matrix. As
a direct consequence, these calibration algorithms fail in one of their
primary mission: they are very poor market risk visualization tools. The
Forward rates covariance matrix plays an increasingly important role in
exotic interest rate derivatives modelling and there is a need for a
calibration algorithm that allows the retrieval of a maximum amount of
covariance information from the market.

In the Libor market model, we write Swaps as baskets of Forwards. As already
observed by \cite{Rebo98} among others, the weights in this decomposition
are empirically very stable. In section two, we show that this key empirical
fact is indeed \textit{accurately reproduced by the model}. We then show
that the drift term coming from the change of measure between the forward
and the swap martingale measures can be neglected in the computation of the
Swaption price, thus allowing these options to be priced using the lognormal
approximations first detailed in \cite{Huyn94} and \cite{Musi97}. In
particular, this will allow us to reduce the problem of pricing Swaptions in
the Libor market model to that of pricing Swaptions in a multidimensional
\cite{Blac73}\ lognormal model. Section three is then focused on finding a
good pricing approximation for basket calls in this generic model. We derive
a simple yet very precise formula where the first term is computed as the
usual \cite{Blac73} price with an appropriate variance and the second term
can be interpreted as approximating the expected value of the tracking error
obtained when hedging with the approximate volatility.

Besides its radical numerical performance compared to Monte-Carlo methods,
the formula we obtain has the advantage of expressing the price of a basket
option in terms of a \cite{Blac73} covariance that is a linear form in the
underlying covariance matrix. This sets the multidimensional model
calibration problem as that of finding a positive semidefinite (covariance)
matrix that satisfies a certain number of linear constraints, in other
words, the calibration becomes a \textit{semidefinite program}. Recent
advances in optimization (see \cite{Nest94} or \cite{Vand96}) have led to
algorithms which solve these problems with a complexity that is comparable
to that of linear programs (see \cite{Nest98}). This means that the general
multidimensional market covariance calibration problem can be solved very
efficiently.

The basket option representation was used in \cite{elka92} where
Swaptions were written as Bond Put options in the Linear Gauss
Markov affine model. \cite{Rebo98} and \cite{Rebo99} detail their
decomposition as baskets of Forwards in the Libor market model. In
parallel results, \cite{Brac00} used semidefinite programming and
the order zero lognormal approximation to study the impact of the
model dimension on Bermudan Swaptions pricing. They rely on
simulation results dating back to \cite{Huyn94}, \cite{Musi97} or
lately \cite{Brac99} in an equity framework to justify the
lognormal volatility approximation of the swap process and they
neglect the change of measure. A big step in the same direction
had also been made by \cite{Rebo99} where the calibration problem
was reparameterized on a hypersphere. However, because it did not
recognize the convexity of the problem, this last method could not
solve the key numerical issue. In recent works, \cite{Sing01}
studied the effect of zero-coupon dynamics degeneracy on Swaption
pricing in an affine term structure model while \cite{JuCF02} use
a Taylor expansion of the characteristic function to derive basket
and Asian option approximations.

This paper is organized around three contributions:

\begin{itemize}
\item  In section two, we detail the basket decomposition of Swaps and
recall some important results on the market model of interest rates. We show
that the weight's volatility and the contribution of the forward vs. swap
martingale measure change can be neglected when pricing Swaptions in that
model.

\item  In section three, we justify the classical lognormal basket option
pricing approximation and compute additional terms in the price expansion.
We also study the implications in terms of hedging and the method's
precision in practice.

\item  In section four, we explicit the general calibration problem
formulation and discuss its numerical performance versus the classical
methods. We specifically focus on the rank issue and its implications in
derivatives pricing. We show how the calibration result can be stabilized in
the spirit of \cite{Cont01} to reduce hedging transaction costs.
\end{itemize}

Numerical instability has a direct cost in both unnecessary hedging
portfolio rebalancing and poor risk modelling. By reducing the amount of
numerical noise in the daily recalibration process and improving the
reliability of risk-management computations, we hope these methods will
significantly reduce hedging costs.

\section{Interest rate market dynamics}

\subsection{Zero coupon bonds and the absence of arbitrage}

We begin here by quickly recalling the construction of the Libor Market
Model along the lines of \cite{BGM97}. We note $B(t,T)$\ the discount
factors (or Zero Coupon bonds) which represent the price in $t$ of one euro
paid at time $T$. We note $\beta _{T}$ the value at time $T$ of one euro
invested in the savings account at $t$ (today) and continuously compounded
with rate $r_{s}$. We have $\beta _{T}=\exp \left(
\int_{t}^{T}r_{s}ds\right) $. As in \cite{Heat92}, to preclude arbitrage
between $\beta _{T}$ and an investment in the Z.C. we impose:
\begin{equation}
B(t,T)=E_{t}^{\mathbf{Q}}\left[ \exp \left( -\int_{t}^{T}r_{s}ds\right) %
\right]  \label{zcprix}
\end{equation}
for some measure $\mathbf{Q}$. In what follows, we will use the Musiela
parametrization of the \cite{Heat92}\ setup and the fundamental rate $%
r(t,\theta )$ will be the continuously compounded instantaneous
forward rate at time $t$, with duration $\theta $. We suppose that
the zero coupon bonds follow a diffusion process driven by a $d$
dimensional $\mathbf{Q}-$Brownian motion $W=\{W_{t},t\geq 0\}$ and
because of the arbitrage argument in (\ref{zcprix}), we know that
the drift term of this diffusion must be equal to $r_{s}$, hence
we can write the zero coupon dynamics as:
\begin{equation}
\frac{dB(s,T)}{B(s,T)}=r_{s}ds+\sigma ^{B}(s,T-s)dW_{s}  \label{ZCdynamics}
\end{equation}
where for all $\theta \geq 0$ the zero-coupon bond volatility process $%
\{\sigma ^{B}(t,\theta );\theta \geq 0\}$ is $F_{t}$-adapted with values in $%
\reals^{d}$. We assume that the function $\theta \longmapsto \sigma
^{B}(t,\theta )$ is absolutely continuous and the derivative $\tau (t,\theta
)=\partial /\partial \theta (\sigma ^{B}(t,\theta ))$ is bounded on $\reals%
^{2}\times \Omega .$ All these processes are defined on the probability
space $(\Omega ,\{F_{t};t\geq 0\},\mathbf{Q})$ where the filtration $%
\{F_{t};t\geq 0\}$ is the $\mathbf{Q}$-augmentation of the natural
filtration generated by the $d$ dimensional Brownian motion $W=\{W_{t},t\geq
0\}$. The absence of arbitrage condition between all zero-coupons and the
savings account then amounts to impose to the process:
\begin{equation}
\frac{B(t,T)}{\beta _{t}}=B(0,T)\exp \left( -\int_{0}^{t}\sigma
^{B}(s,T-s)dW_{s}-\frac{1}{2}\int_{0}^{t}\left| \sigma ^{B}(s,T-s)\right|
^{2}ds\right)  \label{HJMarb}
\end{equation}
to be a martingale under the measure $\mathbf{Q}$ for all $T>0.$

\subsection{Libor rates, Swap rates and the Libor market model}

\subsubsection{Libors and Swaps}

We note $L_{\delta }(t,\theta )$ the forward $\delta $-Libor rate, defined
by:
\begin{equation*}
\frac{1}{1+\delta L_{\delta }(t,\theta )}=\frac{B(t,t+\delta
+\theta )}{B(t,t+\theta )}
\end{equation*}
and we note $K(t,T)=L(t,T-t)$ the forward Libor with constant maturity date
(FRA).

A Swap rate is then defined as the fixed rate that zeroes the
present value of a set of periodical exchanges of fixed against
floating coupons on a Libor rate of given maturity at future dates
$T_{i}^{fx}$ and $T_{i}^{fl}$. This means:
\begin{equation*}
swap(t)=\frac{B(t,T^{fl})-B(t,T_{n+1}^{fl})}{%
Level(t)}
\end{equation*}
where, with $cv(T_{i},T_{i+1})$ the coverage (time interval)
between $T_{i}$ and $T_{i+1}$ computed with the appropriate basis
(different for the floating and fixed legs) and $B(t,T_{i}^{fl})$
the discount factor with maturity $T_{i}^{fl}$, we have defined
$Level(t)$ as the average of the discount factors for the fixed
calendar of the Swap weighted by their associated coverage:
$Level(t)=
\sum_{i=i_{T}}^{n}cv(T_{i}^{fx},T_{i+1}^{fx})B(t,T_{i}^{fx})$.
Here $T_{i}^{fl}$ is the calendar for the floating leg of the swap
and $T_{i}^{fx}$ is the calendar for the fixed leg (the notation
is there to highlight the fact that they don't match in general).
In a representation that will be critically important in the
pricing approximations that follow, we remark that we can write
the Swaps as baskets of Forward Libors (see for ex.
\cite{Rebo98}).

\begin{lemma}
We can write the Swap with floating leg $T^{fl},\ldots,T_{N}^{fl}$
as a basket of Forwards:
\begin{equation}
swap(t)=\sum_{i=i_{T}}^{n}\omega _{i}(t)K(t,T_{i}^{fl})\text{ \ \
where \ }\omega _{i}(t)=\frac{
cv(T_{i}^{fl},T_{i+1}^{fl})B(t,T_{i+1}^{fl})}{%
Level(t)}  \label{BasketForwards}
\end{equation}
with $T_{i_{T}}=T$ and $0\leq \omega _{i}(t)\leq 1$.
\end{lemma}

\begin{proof}
With $B(t,T_{i}^{fl})=B(t,T_{i+1}^{fl})(1+\delta
K(t,T_{i}^{fl})),$ we have:
\begin{equation*}
swap(t)=\frac{
\sum_{i=i_{T}}^{n}cv(T_{i}^{fl},T_{i+1}^{fl})B(t,T_{i+1}^{fl})K(t,T_{i}^{fl})%
}{Level(t)}
\end{equation*}
which is the desired representation. As the corresponding forward Libor
rates are positive, we have $B(t,T_{i+1})\leq B(t,T_{i})\leq B(t,T_{i-1})$ \
for \ $i\in \lbrack i_{T}+1,N-1]$ hence $0\leq \omega _{i}(t)\leq 1$, i.e.
the weights are positive and bounded by one.
\end{proof}

As we will see below, the weights $\omega _{i}(t)$ prove to have
very little variance compared to their respective FRA (see
\cite{Rebo98} among others). This approximation of Swaps as
baskets of Forwards with constant coefficients is the key factor
behind the Swaption pricing methods that we detail here.

\subsubsection{The Libor market model}

As Libor rates and Swaps were gaining importance as the fundamental
variables on which the market activity was concentrated, a set of options
was created on these market rates: the Caps and Swaptions. Adapting the
common practice taken from equity markets and the \cite{Blac73} framework,
market operators looked for a model that would set the dynamics of the
Libors or the Swaps as lognormal processes. Intuitively, the lognormal
assumption on prices can be justified as the effect of a central limit
theorem on returns because the prices are seen as driven by a sequence of
independent shocks on returns. That same reasoning cannot be applied to
justify the lognormality of Libor or Swap rates, which are rates of return
themselves. The key justification behind this assumption must then probably
be found in the legibility and familiarity of the pricing formulas that are
obtained: the market quotes the options on Libors and Swaps in terms of
their \cite{Blac76} volatility by habit, it then naturally tries to model
the dynamics of these rates as lognormal.

Everything works fine when one looks at these prices and processes
individually, however some major difficulties arise when one tries to define
yield curve dynamics that jointly reproduce the lognormality of Libors and
Swaps. In fact, it is not possible to find arbitrage free dynamics \`{a} la
\cite{Heat92} that make both Swaps and Libors lognormal under the
appropriate forward measures (see \cite{Musi97} or \cite{Jams97} for an
extensive discussion of this). Here we choose to adopt the \cite{Heat92}
model structure defined in \cite{BGM97} (see also \cite{Milt95} or \cite
{Sand97}) where the Libor rates are specified as lognormal under the
appropriate forward measures but we will see in a last section that for the
purpose of pricing options on Swaps, one can in fact approximate the swap by
a lognormal diffusion. Hence in a very reassuring conclusion on the model,
observed empirically in \cite{Brac99}, we notice that it is in fact possible
to specify \cite{Heat92} dynamics that are reasonably close to the market
practice, i.e. lognormal on Forwards and close (in a sense that will be made
clear later) to lognormal on Swaps. In particular, we verify that the key
property behind this approximation, namely the stability of the weights $%
\omega _{i}(t),$ is indeed \textit{accurately reproduced} by the Libor
market model.

The model starts from the key assumption that for a given maturity $\delta $
(for ex. 3 months) the associated forward Libor rate process has a
log-normal volatility structure:
\begin{equation}
dL(t,\theta )=(...)dt+L(t,\theta )\gamma (t,\theta )dW_{t}  \label{vol def1}
\end{equation}
where the deterministic function $\gamma :\reals_{+}^{2}\longmapsto \reals%
_{+}^{d}$ is bounded by some $\bar{\gamma}\in \reals_{+}$ and piecewise
continuous. As for all \cite{Heat92} based models, these dynamics are fully
specified by the definition of the volatility structure and the forward
curve today. With that in mind, we derive the appropriate zero-coupon
volatility expression. Using the Ito formula combined with (\ref{HJMarb}) we
get as in \cite{BGM97}:
\begin{align*}
dL(t,\theta )& =\left( \frac{\partial L(t,\theta )}{\partial \theta }+\frac{%
\left( 1+\delta L(t,\theta )\right) }{\delta }\sigma ^{B}(t,\theta +\delta
)(\sigma ^{B}(t,\theta +\delta )-\sigma ^{B}(t,\theta ))\right) dt \\
& +\frac{1}{\delta }\left( 1+\delta L(t,\theta )\right) (\sigma
^{B}(t,\theta +\delta )-\sigma ^{B}(t,\theta ))dW_{t}
\end{align*}
Then to get the right volatility structure we have to impose in (\ref
{ZCdynamics}):
\begin{equation}
\sigma ^{B}(t,\theta +\delta )-\sigma ^{B}(t,\theta )=\frac{\delta
L(t,\theta )}{1+\delta L(t,\theta )}\gamma (t,\theta )  \label{vol def2}
\end{equation}
The Libor process becomes:
\begin{equation*}
dL(t,\theta )=\left( \frac{\partial }{\partial \theta }L(t,\theta )+\gamma
(t,\theta )\sigma ^{B}(t,\theta +\delta )L(t,\theta )\right) dt+L(t,\theta
)\gamma (t,\theta )dW_{t}
\end{equation*}
As in \cite{Musi97}, we set $\sigma ^{B}(t,\theta )=0$ for all $\theta \in
\lbrack 0,\delta \lbrack $ and we get, together with the recurrence relation
(\ref{vol def2}) and for $\theta \geq \delta $ :
\begin{equation}
\sigma ^{B}(t,\theta )=\sum_{k=1}^{\lfloor \delta ^{-1}\theta \rfloor }\frac{%
\delta L(t,\theta -k\delta )}{1+\delta L(t,\theta -k\delta )}\gamma
(t,\theta -k\delta )  \label{vol recurence}
\end{equation}
With the volatility of the zero coupon defined above and the value of the
forward curve today, we have fully specified the yield curve dynamics.

\subsection{Interest rate options: Caps and Swaptions}

\subsubsection{Caps}

Let us note again $\beta (t),$ the value of the savings account. In a
forward Cap on principal 1 settled in arrears at times $T_{j}$, $j=1,...,n,$
the cash-flows are $(L(T_{j-1},0)-K)^{+}\delta $ paid at time $T_{j}$. The
price of the Cap at time t is then computed as:
\begin{equation}
Cap_{t}=\sum_{1}^{n}E_{t}^{\mathbf{Q}}\left[ \frac{\beta _{t}}{\beta _{T_{j}}%
}\left( L(T_{j-1},0)-k\right) ^{+}\delta \right]  \notag
\end{equation}

\subsubsection{Swaptions}

To simplify the notations, we will consider that the calendars described
above for the floating and the fixed legs of the swap are set by $%
T_{i}^{fl}=i\delta $ and $T_{i}^{fx}=ib\delta $, in the common
case where the fixed coverage is a multiple of the floating
coverage (for ex. quarterly floating leg, annual fixed leg). For
simplicity, we will note the coverage function for the fixed leg
of the swap as a function of the floating dates, allowing the
floating dates to be used as reference in the entire swap
definition. From now on $(T_{i})_{i\in \lbrack
1,N]}=(T_{i}^{fl})_{i\in \lbrack 1,N]}$ and we define the coverage
function for the fixed leg as $c_i\delta
=1_{\{i\func{mod}b=0\}}b\delta .$ We set $i_{T}=\lfloor \delta
^{-1}T\rfloor $. Using these simplified notations the Swap in
(\ref{BasketForwards}) becomes:
\begin{equation*}
swap(t)=\frac{B(t,T)-B(t,T_{N+1})}{Level(t)}\text{ \ \ with \ \
}Level(t)=\sum_{i=i_{T}}^{N}\delta c_i B(t,T_{i+1})
\end{equation*}
The price of a payer Swaption with maturity $T$ and strike $k$, written on
this swap is then given at time $t\leq T$ \ by:
\begin{equation}
Swaption_{t}=E_{t}^{Q}\left[ \sum_{i=i_{T}}^{N}\frac{\beta
(t)}{\beta (T_{i+1})}c_i\delta \left( swap(T)-k\right) ^{+}\right]
\label{SwaptionPriceDef}
\end{equation}
The expression above computes the price of the Swaption as the sum\ of the
corresponding Swaplet prices. Because a Caplet is an option on a one period
Swap, Caplet and Swaption prices can be computed in the same fashion. In the
two sections that follow, we show how to rewrite this pricing expression to
describe the Swaption (and the Caplet) as a basket option.

\subsection{Caps and Swaptions in the Libor market model}

\subsubsection{Caps and the forward martingale measure}

With the Cap price computed as:
\begin{equation*}
Cap_{t}=\sum_{j=1}^{n}B(t,T_{j})E_{t}^{T_{j}}\left[ \left(
L(T_{j-1},0)-K\right) ^{+}\delta \right]
\end{equation*}
where $E^{T_{j}}$ is the expectation under the forward martingale measure $%
\mathbf{Q}_{Tj}$ defined by:
\begin{equation*}
\frac{d\mathbf{Q}_{Tj}}{d\mathbf{Q}}=[B(0,T)\beta _{T}]^{-1}=\varepsilon
_{T}(\sigma ^{B}(\cdot ,T_{j}-\cdot ))
\end{equation*}
where we have noted $\varepsilon _{T}(\cdot )$ the exponential martingale
defined by:
\begin{equation*}
\varepsilon _{T}(\sigma ^{B}(\cdot ,T_{j}-\cdot ))=\exp \left(
\int_{0}^{T_{j}}\sigma ^{B}(s,T_{j}-s)dW_{s}-\frac{1}{2}\int_{0}^{T_{j}}%
\left\| \sigma ^{B}(s,T_{j}-s)\right\| ^{2}ds\right)
\end{equation*}
Let us now define the forward Libor process (or FRA) dynamics, the
underlying $K(t,T)=L(t,T-t)$ of the Caplet paid at time $T+\delta $, which
is given in the Libor market model setup in (\ref{vol def1}) by:
\begin{equation*}
dK(t,T)=\gamma (t,T-t)K(t,T)\left[ \sigma ^{B}(t,T-t+\delta )dt+dW_{t}\right]
\end{equation*}
or again:
\begin{equation}
dK(t,T)=\gamma (t,T-t)K(t,T)dW_{t}^{T+\delta }  \label{libor-forward-proc}
\end{equation}
hence $K(t,T)$ is lognormally distributed under $P_{T+\delta }$.
Here and in what follows, we note $V_{t,T}$ the cumulative
variance from $t$ to $T$ and the pricing of Caplets can be done
using the \cite{Blac76} formula with $V_{t,T}$ equal to:
\begin{equation*}
V_{t,T}=\int_{t}^{T}\left\| \gamma (s,T-s)\right\| ^{2}ds
\end{equation*}
Let us note that the Caplet variance used in the \cite{Blac76} pricing
formula is a linear form in the covariance. Recovering the same kind of
result in the Swaption pricing approximation will be the key to the
calibration algorithm design.

\subsubsection{Swaptions and the forward swap martingale measure}

In (\ref{SwaptionPriceDef}) the price of a payer Swaption is
computed as the sum\ of the corresponding Swaplet prices, which is
not the most appropriate format for pricing purposes. Using a
change of equivalent probability measure, we now find another
expression that is more suitable for our analysis. As in
\cite{Musi97}, we can define the \textit{forward swap martingale
probability measure} $\mathbf{Q}^{S}$ equivalent to
$\mathbf{Q}^{T}$, with:
\begin{align*}
\frac{d\mathbf{Q}^{S}}{d\mathbf{Q}^{T}}|_{t}& =\frac{
\sum_{i=i_{T}}^{N}c_i\beta (T)/\beta
(T_{i+1})}{E_{t}^{Q_{T}}\left[
\sum_{i=i_{T}}^{N}c_i\beta (T)/\beta (T_{i+1})\right] } \\
& =B(t,T)\beta (T)\sum_{i=i_{T}}^{N}\frac{\delta c_i\beta
^{-1}(T_{i+1})}{Level(t)}
\end{align*}
This equivalent probability measure corresponds to the choice of
the ratio of the level payment over the savings account as a
numeraire and the above relative bond prices are
$\mathbf{Q}^{T}-$local martingale. The change of measure is
identified with an exponential (local) $\mathbf{Q}^{T}-$
martingale and we define the process $h_{t}$ such that:
\begin{equation*}
\varepsilon _{T_{N}}(h_{\cdot })=B(t,T)\beta
(T)\frac{\sum_{i=i_{T}}^{N} \delta c_i\beta
^{-1}(T_{i+1})}{Level(t)}
\end{equation*}
which imposes:
\begin{equation}
h_{t}=\sum_{i=i_{T}}^{N}\frac{\delta c_iB(t,T_{i+1})}{Level(t)}%
\left( \sum_{j=i_{T}}^{i}\frac{\delta K(t,T_{j})}{1+\delta K(t,T_{j})}\gamma
(t,T_{j}-t)\right)
\end{equation}
and because the volatility is bounded, we verify that $\varepsilon
_{T_{N}}(h_{\cdot })$ is in fact a martingale. Again as in \cite{Musi97} we
can apply Girsanov's theorem to show that the process:
\begin{equation}
dW_{t}^{S}=dW_{t}^{T}+\sum_{i=i_{T}}^{N}\left( \frac{\delta
c_iB(t,T_{i+1})}{Level(t)}\sum_{j=i_{T}}^{i}\frac{\delta
K(t,T_{j})}{1+\delta K(t,T_{j})}\gamma (t,T_{j}-t)\right) dt
\label{LVLbrownian}
\end{equation}
is a $\mathbf{Q}^{S}$-Brownian motion.

\begin{lemma}
We can rewrite the Swaption price as:
\begin{equation}
Swaption_{t}=Level(t)E_{t}^{Q_{S}}\left[ \left( swap(T)-k\right)
^{+}\right]
\end{equation}
\textit{where the swap rate is a martingale under the new probability measure%
} $\mathbf{Q}^{S}$.
\end{lemma}

\begin{proof}
The pricing formula is a direct consequence of the change of
measure above and because the Swap is defined by the ratio of a
difference of zero-coupon prices over the level payment, it is a
(local) martingale under the new probability measure
$\mathbf{Q}^{S}$ (below, we will see that the swap rate is in fact
a $\mathbf{Q}^{S}-$martingale).
\end{proof}

This change of measure first detailed by \cite{Jams97}, allows to price
Swaptions as classical Call options on a swap, under an appropriate measure.

\subsection{Swap dynamics}

We now study the dynamics of the swap rate under the
$\mathbf{Q}^{S}$ probability, looking first for an appropriate
representation of the volatility function using the ''basket of
forwards'' decomposition $swap(t)=\sum_{i=i_{T}}^{N}\omega
_{i}(t)K(t,T_{i})$ detailed in (\ref {BasketForwards}).

\begin{lemma}
The weights $\omega _{k}(s)$ in the swap decomposition follow:
\begin{equation*}
d\omega _{k}(s)=\omega _{k}(s)\sum_{i=i_{T}}^{N}\omega
_{i}(s)\left( \sigma ^{B}(s,T_{k+1}-s)-\sigma
^{B}(s,T_{i+1}-s)\right) dW_{s}^{S}
\end{equation*}
\end{lemma}

\begin{proof}
As the ratio of a zero coupon bond on the level payment and by
construction of $\mathbf{Q}^{S}$, the weights $\omega _{i}(t)$
must be $\mathbf{Q}^{S}-$martingales (they are positive bounded).
Using the forward zero-coupon dynamics, we then get:
\begin{eqnarray*}
d\left( \frac{B(s,T_{k})}{Level(t)}\right) &=&(...)ds+\frac{
B(t,T_{k})}{Level(t)}\sigma ^{B}(s,T_{k}-s)dW_{s}^{T} \\
&&-\frac{B(t,T_{k})}{Level(t)}\sum_{i=i_{T}}^{N}\frac{\delta
c_iB(t,T_{i+1})}{Level(t)}\sigma ^{B}(s,T_{i+1}-s)dW_{s}^{T}
\end{eqnarray*}
where $W_{s}^{T}$ is a $\mathbf{Q}^{T}$-Brownian motion.
\end{proof}

We then use this result to decompose the Swap volatility as the
sum of the weights volatility term and a term that mimics a
\textit{basket volatility} (the volatility of a basket with
constant coefficients). We write the swap volatility as:
\begin{equation}
dswap(s)=\sum_{i=i_{T}}^{N}\omega _{i}(s)K(s,T_{i})\left( \gamma
(s,T_{i}-s)+\eta (s,T_{i})\right) dW_{s}^{S} \label{swapdynamics}
\end{equation}
where the \textit{basket volatility} term and the \textit{weight's residual
contribution} are given by:
\begin{equation*}
\sum_{i=i_{T}}^{N}\omega _{i}(s)K(s,T_{i})\gamma (s,T_{i}-s)\text{ \ \ and \
\ }\eta (s,T_{i})=\left( \sigma ^{B}(s,T_{i+1}-s)-\sum_{j=i_{T}}^{N}\omega
_{j}(s)\sigma ^{B}(s,T_{j+1}-s)\right)
\end{equation*}
Again, the empirical stability of the weights $\omega _{i}(t)$ is
the key fact at the origin of the Swaption pricing approximations
that will follow and one of our goals below will be to show that
this stability is \textit{accurately reproduced by the model}.

\subsection{The forward Libors under the forward Swap measure}

We study here the dynamics of the forward Libors under the forward
Swap measure. For purely technical purposes, we start by bounding
under $\mathbf{Q}^{S}$ the variance of the forward rates
$K(s,T_{k})$, this will allow us to bound the contribution of the
weights to the total swap variance.

\begin{lemma}
With $m>1$, we can bound the $L^{2}$ norm of $K(u,T_{k})$ under
$\mathbf{Q}^{S}$ by:
\begin{equation}
E[K(s,T_{k})^{m}]\leq K(t,T_{k})^{m}M_{m}^{m}(s)  \label{forwardL2bound}
\end{equation}

where $M_{m}(s)=\exp \left( (s-t)\left(
m\bar{\gamma}^{2}/2+m\bar{\gamma}^{2}\delta (N-i_{T})\right)
\right) $.
\end{lemma}

\begin{proof}
Using (\ref{LVLbrownian}) we can write:
\begin{equation*}
K(s,T_{k})=K(t,T_{k})\exp \left( \int_{t}^{s}\gamma
(u,T_{k}-u)dW_{u}^{S}+\int_{t}^{s}\alpha (u,T_{k})\gamma
(u,T_{k}-u)du\right)
\end{equation*}
where
\begin{equation*}
\alpha (s,T_{k})=-\sum_{i=i_{T}}^{N}\omega _{i}(s)\left(
\sum_{j=i_{T}}^{i}\phi _{j}(s)\gamma (s,T_{j}-s)\right)
+\sum_{i=i_{T}}^{k}\phi _{i}(s)\gamma (s,T_{i}-s)
\end{equation*}
with $\phi _{i}(t)=\delta K(s,T_{i})/(1+\delta K(s,T_{i})).$ The
corresponding forward Libor rates are positive and we have $0\leq \phi
_{i}(t)\leq 1$ and as in \cite{BGM97} remark 2.3, we can bound the Forwards
by a lognormal process:
\begin{equation*}
K(s,T_{k})\leq K(t,T_{k})\exp \left( \int_{t}^{s}\gamma
(u,T_{k}-u)dW_{u}^{S}+\int_{t}^{s}\bar{\alpha}(u,T_{k})du\right)
\text{ \ for }s\in \lbrack t,T]
\end{equation*}
where we can use a convexity inequality on the norm $\left\|
.\right\|^{2}$ to obtain:
\begin{equation*}
\left\| \sum_{i=i_{T}}^{N}\omega _{i}(s)\left( \sum_{j=i_{T}}^{i}\phi
_{j}(s)\gamma (s,T_{j}-s)\gamma (s,T_{k}-s)\right) \right\| ^{2}\leq \delta
^{2}(N-i_{T})^{2}\bar{\gamma}^{4}
\end{equation*}
because$\left\| \sum_{i=i_{T}}^{k}\phi _{i}(t)\gamma (s,T_{i}-s)\gamma
(s,T_{k}-s)\right\| ^{2}\leq \delta ^{2}(k-i_{T})^{2}\bar{\gamma}^{4}$,
hence $\bar{\alpha}(s,T_{k})=\delta (N-i_{T})\bar{\gamma}^{2}$ which shows
the desired result.
\end{proof}

We now use this bound to study the impact of the weights $\omega _{i}(t)$ in
the swap volatility decomposition.

\subsection{Swaps as baskets of forwards}

For simplicity, in what follows we will suppose that
$T_{i}^{fl}=T_{i}^{fx}$ and hence $b=1$. The Swaption pricing
formula that will be derived in section (\ref{s-basket-approx})
relies on two fundamental approximations:

\begin{itemize}
\item  The weights $\omega _{i}(s)$ for $s\in \lbrack t,T]$
(which are $\mathbf{Q}^{S}$-martingales) will be approximated by
their value today $\omega _{i}(t)$.

\item  We will neglect the change of measure between the forward martingale
measures $\mathbf{Q}^{T}$ to $\mathbf{Q}^{T_{N+1}}$ and the
forward Swap martingale measure $\mathbf{Q}^{S}$.
\end{itemize}

In this section, we study the impact of these approximations and
try to quantify the pricing error they induce. The consequences of
the first approximation are studied in lemma
(\ref{lemma-weights}), while proposition (\ref{prop-swap-proc})
describes the impact of the second. The low weight volatility is
steadily observed in practice, besides \cite{Rebo98}, this has
been studied by \cite{Hamy99} of which we report here, with the
author's permission, a sample of summary statistics. The table
below details the $vol(FRA)/vol(weights)$ ratio in various
markets, computed using the standard quadratic variation estimator
with exponentially decaying weights (market data courtesy of
BNP-Paribas London):\
\begin{gather*}
\begin{array}{lcccccc}
\text{Currency} & \text{{\small USD}} & \text{{\small USD}} & \text{{\small %
GBP}} & \text{{\small GBP}} & \text{{\small EUR}} & \text{{\small EUR}} \\
\text{swap} & 2Y & 5Y & 2Y & 5Y & 2Y & 5Y \\
\text{Min ratio} & 712 & 842 & 885 & 981 & 148 & 333 \\
\text{Max ratio} & 7629 & 7927 & 6575 & 3473 & 5006 & 4322 \\
\text{Variance} & .023 & .020 & .017 & .007 & .005 & .004
\end{array}
\\
\text{Sample ratio of volatility between weights and corresponding
forwards.}
\end{gather*}
Here \textit{Min ratio} and \textit{Max ratio} are the minimum
(resp. maximum) volatility ratio among the weights of a particular
swap. We see that in this sample, the volatility of the weights is
always several orders of magnitude lower than the volatility of
the corresponding forward. Also, the weights in
(\ref{BasketForwards}) are positive, monotone and sum to one,
cancelling the first order error terms, hence if the Forward rate
curve is flat ($K(s,T_{i})=K(s,T_{j})$ for $i,j=i_{T},...,N$) we
have:
\begin{equation*}
\sum_{i=i_{T}}^{N}\omega _{i}(s)K(s,T_{i})\eta
(s,T_{i})=K(s,T_{i})\sum_{i=i_{T}}^{N}\omega _{i}(s)\left(
\sigma^{B}(s,T_{i}-s)-\sum_{j=i_{T}}^{N}\omega
_{j}(s)\sigma^{B}(s,T_{j}-s)\right) =0
\end{equation*}
In light of this, we will study the size of the weights'
contribution to the swap volatility in terms of the slope of the
Forward rate curve within the maturity range of the swap's
floating leg. In particular, we can write the weight's part in the
Swap's volatility as:
\begin{equation}
E^{S}\left[ \left\| \sum_{i=i_{T}}^{N}\omega _{i}(s)K(s,T_{i})\eta
(s,T_{i})\right\| ^{2}\right] =E^{S}\left[ \left\|
\sum_{i=i_{T}}^{N}\omega _{i}(s)\left( K(s,T_{i})-swap(s)\right)
\eta (s,T_{i})\right\| ^{2}\right]  \label{diffequality}
\end{equation}
which sets the weight's contribution as the average product of a
difference of Forwards with a difference of ZC\ bond volatilities
and we can expect this later term to be negligible relative to the
basket volatility term in (\ref {swapdynamics}), in accordance
with the empirical evidence. Because the payoff of the Call
options under consideration are Lipschitz, we will approximate the
Swap and forward Libor dynamics in $L^{2}$ under the
$\mathbf{Q}^{S}$ swap martingale measure.

We now detail some basic properties of the weights $\omega
_{i}(s)$. We note $\left\| \cdot \right\| _{n}=\left( E^{S}\left[
\left\| \cdot \right\| ^{n}\right] \right) ^{1/n}$, the $L^{n}$
norm.

\begin{lemma}
The weights $\omega _{i}(s)$ defined in (\ref{BasketForwards}) are bounded
above with:
\begin{equation*}
\omega _{i}(s)\leq \frac{1}{N-i_{T}}+\delta swap(s)
\end{equation*}
and satisfy $\left\| \omega _{i}(s)\right\| _{n}\leq \omega
_{i}(t)$\ for $s\in \lbrack t,T].$
\end{lemma}

\begin{proof}
Because the weights $\omega _{i}(s)$ satisfy $\sum_{i=i_{T}}^{N}\omega
_{i}(t)=1$, $0\leq \omega _{i}(t)\leq 1$ and are decreasing with $i$ because
the Forward rates are always positive. With:
\begin{equation*}
\left| \omega _{j}(s)-\omega _{i}(s)\right| \leq \delta
swap(s)\text{ \ for }i,j=i_{T},...,N
\end{equation*}
we get:
\begin{equation*}
\omega _{i}(s)\leq \frac{1}{N-i_{T}}+\delta swap(s)\text{ \ for }
s\in \lbrack t,T]
\end{equation*}
and $\left\| \omega _{i}(s)\right\| _{n}\leq \left\| \omega
_{i}(s)\right\| _{1}=\omega _{i}(t)$, for $s\in \lbrack t,T]$ and
$n\geq 1$, because the weights are positive
$\mathbf{Q}^{S}-$martingales.
\end{proof}

The next result provides a bound on the variance contribution of the weights
inside the Swap rate volatility.

\begin{lemma}
The $L^{2}$ norm of the weight's contribution in the swap volatility (\ref
{swapdynamics}) is bounded by:
\begin{eqnarray}
&&E^{S}\left[ \left\| \sum_{i=i_{T}}^{N}\omega
_{i}(s)K(s,T_{i})\eta
(s,T_{i})\right\| ^{2}\right]  \label{contribneglect} \\
&\leq &\max_{j}\left\| \left( K(s,T_{j})-swap(s)\right) \right\|
_{8}^{2}M_{4}^{2}\bar{\gamma}^{2}\delta ^{2}max_{j\in \lbrack
i_{T},N]}K(t,T_{j})^{2}(N-i_{T})^{2}  \notag
\end{eqnarray}
\label{lemma-weights}
\end{lemma}

\begin{proof}
Let us note again $swap(s)=\sum_{i=i_{T}}^{N}\omega
_{i}(s)K(s,T_{i}),$ the swap rate, which we see here as the
average level of the Forward rate curve between $T$ and $T_{N}$.
The squared $L^{2}$ norm of the weights' contribution is bounded
above by:
\begin{equation*}
E^{S}\left[ \sum_{i=i_{T}}^{N}\omega _{i}(s)\left\| \left(
K(s,T_{i})-swap(s)\right) \eta (s,T_{i})\right\|^{2}\right]
\end{equation*}
using a convexity inequality with $\sum_{i=i_{T}}^{N}\omega _{i}(t)=1$, $%
0\leq \omega _{i}(t)\leq 1$. To bound $\eta (s,T_{k})$ in this expression,
we use the definition of $\sigma ^{B}(s,T_{k}-s)$ in (\ref{vol recurence})
and the fact that the Forwards $K(s,T_{j})$\ are always positive to get:
\begin{equation*}
E^{S}\left[ \left\| \eta (s,T_{i})\right\| ^{4}\right] \leq
E^{S}\left[ \left\| \sum_{i=i_{T}}^{N}\omega _{i}(s)\left(
\sum_{j=i}^{k}\delta K(s,T_{j})\gamma (s,T_{j}-s)\right) \right\|
^{4}\right]
\end{equation*}
with the convention $\sum_{j=i}^{k}=-\sum_{j=k}^{i}$ if $i>k$. If we recall
that $\gamma (s,T_{k}-s):\reals_{+}^{2}\rightarrow \reals_{+}^{d}$ is a
bounded input parameter with $E\left[ \left\| \gamma (s,T_{k}-s)\right\| ^{2}%
\right] \leq \bar{\gamma}^{2}$, we can use (\ref{forwardL2bound}) and the
previous lemma to get:
\begin{equation*}
E^{S}\left[ \left\| \eta (s,T_{k})\right\| ^{4}\right] \leq M_{4}^{4}\bar{%
\gamma}^{4}\delta ^{4}max_{j\in \lbrack i_{T},N]}K(t,T_{j})^{4}(N-i_{T})^{4}
\end{equation*}
With these bounds we can rewrite the original inequality, using two
successive Cauchy inequalities:
\begin{eqnarray*}
&&E^{S}\left[ \sum_{i=i_{T}}^{N}\omega _{i}(s)\left(
K(s,T_{i})-swap(s)\right) ^{2}\left\| \eta (s,T_{i})\right\|^{2}\right] \\
&\leq &\sum_{i=i_{T}}^{N}\left\| \omega _{i}(s)\right\|
_{4}\left\| \left( K(s,T_{i})-swap(s)\right) \right\|
_{8}^{2}\left\| \eta (s,T_{i})\right\|_{4}^{2} \\
&\leq &\max_{j}\left\| \left( K(s,T_{j})-swap(s)\right) \right\|
_{8}^{2}M_{4}^{2}\bar{\gamma}^{2}\delta ^{2}max_{j\in \lbrack
i_{T},N]}K(t,T_{j})^{2}(N-i_{T})^{2}
\end{eqnarray*}
Which gives the desired result.
\end{proof}

With $\delta K(t,T_{k})\simeq 10^{-2}$ and $\left(
K(s,T_{i})-swap(s)\right) ^{2}\simeq 10^{-3}$ in practice, we
notice that the contribution of the weights to the swap volatility
is several orders of magnitude below that of the basket and we
will neglect it in the Swaptions pricing approximations that
follow. Before detailing the key approximation result, we
introduce some new notations.

\begin{notation}
We define $K^{S}(s,T_{i})$ such that:
\begin{equation*}
dK^{S}(s,T_{i})=K^{S}(s,T_{i})\gamma(s,T_{i}-s)dW_{s}^{S}
\end{equation*}
with $K^{S}(t,T_{i})=K(t,T_{i}).$ We also define the following
residual volatilities:
\begin{equation*}
\xi _{k}(s)=K^{S}(s,T_{k})\gamma(s,T_{k}-s)-\gamma ^{w}(s)
\end{equation*}
with $\gamma ^{w}(s)=\sum_{i=i_{T}}^{N}\omega
_{i}(t)K^{S}(s,T_{i})\gamma (s,T_{k}-s)$.
\end{notation}

We now approximate the Swap rate with a basket of lognormal
martingales parameterized by the Forward rate volatilities $\gamma
(s,T_{k}-s)$ and their initial value $K(t,T_{i}),$ the weights in
this decomposition being equal to $\omega _{i}(t).$

\begin{proposition}
We can replace the Swap process by a basket $Y_{s}$\ of lognormal
martingales weighted by constant coefficients, with:
\begin{eqnarray*}
&&E\left[ \left( \sup_{t\leq s\leq T}\left( swap(s)-Y_{s}\right)
\right) ^{2}\right] \\
&\leq &3\max_{j\in \lbrack i_{T},N]}\left\| \xi _{j}(s)\right\|
_{4}^{2}+3(K^{S}(t,T_{k})\left( N-i_{T}\right) \delta \bar{
\gamma}^{2})^2 \exp \left(2(T-t)\left( \delta
\bar{\gamma}^{2}\left( N-i_{T}\right) +\bar{\gamma}^{2}/2\right) \right) \\
&&+3\max_{j\in \lbrack i_{T},N]}\left\| \left(
K(s,T_{j})-swap(s)\right) \right\| _{8}^{2}M_{4}^{2}\bar{\gamma}
^{2}\delta ^{2}max_{j\in \lbrack
i_{T},N]}K(t,T_{j})^{2}(N-i_{T})^{2}
\end{eqnarray*}
where
\begin{equation*}
dY_{s}=\sum_{i=i_{T}}^{N}\omega _{i}(t)K^{S}(s,T_{i})\gamma
(s,T_{i}-s)dW_{s}^{S}
\end{equation*}
with $Y_{t}=swap(t)$.
\label{prop-swap-proc}
\end{proposition}

\begin{proof}
With the swap rate dynamics computed as in (\ref{swapdynamics}), we get:
\begin{eqnarray*}
d(swap(s)-Y_{s}) &=&\sum_{k=i_{T}}^{N}\left( \omega _{k}(s)-\omega
_{k}(t)\right) K^{S}(s,T_{k})\gamma (s,T_{k}-s)dW_{s}^{S} \\
&&+\sum_{k=i_{T}}^{N}\omega _{k}(s)\left(
K(s,T_{k})-K^{S}(s,T_{k})\right)
\gamma (s,T_{k}-s)dW_{s}^{S} \\
&&+\sum_{k=i_{T}}^{N}\omega _{k}(s)K(s,T_{k})\eta
(s,T_{k})dW_{s}^{S}
\end{eqnarray*}
We can bound the norm of the last term in this decomposition using
the result in (\ref{contribneglect}). If we look at the first term
and note $\Delta _{k,s}=K(s,T_{k})-K^{S}(s,T_{k})$ with $\Delta
_{k,t}=0$ we have:
\begin{eqnarray*}
d\Delta _{k,s} &=&\Delta _{k,s}\left( \sum_{i=i_{T}}^{N}\omega _{i}(s)\left(
\sigma ^{B}(s,T_{k}-s)-\sigma ^{B}(s,T_{i}-s)\right) \gamma
(s,T_{k}-s)\right) \\
&&+K^{S}(s,T_{k})\left( \sum_{i=i_{T}}^{N}\omega _{i}(s)\left(
\sigma
^{B}(s,T_{k}-s)-\sigma ^{B}(s,T_{i}-s)\right) \gamma (s,T_{k}-s)\right) ds \\
&&+\Delta _{k,s}\gamma (s,T_{k}-s)dW_{s}^{S}
\end{eqnarray*}
hence:
\begin{equation*}
\Delta _{k,T}=K^{S}(T,T_{k})\int_{t}^{T}\left( \mu _{k,s}\exp
\left( \int_{t}^{s}\mu _{k,u}du\right) ds\right)
\end{equation*}
where
\begin{equation*}
\mu _{k,s}=\sum_{i=i_{T}}^{N}\omega _{i}(s)\left( \sigma
^{B}(s,T_{k}-s)-\sigma ^{B}(s,T_{i}-s)\right) \gamma (s,T_{k}-s)
\end{equation*}
With $\left\| \mu _{k,s}\right\| _{2}\leq \left( N-i_{T}\right)
\delta \bar{ \gamma}^{2}$ we can bound the norm of $\Delta _{k,T}$
by:
\begin{equation*}
\left\| \Delta _{k,T}\right\| _{2}\leq K(t,T_{k}) \left(
N-i_{T}\right) \delta \bar{ \gamma}^{2} \exp \left( (T-t)\left(
\delta \bar{\gamma}^{2}\left( N-i_{T}\right)
+\bar{\gamma}^{2}/2\right) \right)
\end{equation*}
Focusing on the second term, as in (\ref{diffequality}) with this
time $\sum_{i=i_{T}}^{N}\omega _{i}(s)-\omega _{i}(t)=0$ and $\xi
_{k}(s)=K^{S}(s,T_{k})\gamma (s,T_{k}-s)-\gamma ^{w}(s)$, we can
write:
\begin{equation*}
\left\| \sum_{k=i_{T}}^{N}\left( \omega _{k}(s)-\omega
_{k}(t)\right) K^{S}(s,T_{k})\gamma (s,T_{k}-s)\right\|
_{2}^{2}\leq \max_{j\in \lbrack i_{T},N]}\left\| \xi
_{j}(s)\right\| _{4}^{2}
\end{equation*}
The bound obtained\ is a function of the norm of the residual
volatilities $\left\| \xi _{i}(s)\right\| _{4}^{2}$ and of the
spread term $\left\| \left( K(s,T_{i})-swap(s)\right) \right\|
_{8}^{2}$. We conclude using Doob's inequality.
\end{proof}

The term $\left\| \xi _{i}(s)\right\| _{4}^{2}$ is equivalent to
the variance contribution of the second factor of the covariance
matrix and $\left\| \left( K(s,T_{i})-swap(s)\right) \right\|
_{8}^{2}$ is a spread of rates, so we neglect both terms relative
to the central volatility $\gamma ^{w}(s)$ and we consider the
Swaption as an option on the basket $Y_{s}$. We notice that
because we approximate one martingale by another, the error is in
fact uniformly bounded in $L^{2}$. Because of these properties and
the fact that the option's payoff is Lipschitz, in the Swaption
price approximations that follow, we will be treating the
\textbf{Swaption as an option on a basket of lognormal Forwards}.

\section{Basket price approximation}
\label{s-basket-approx}

Basket options, i.e. options on a basket of goods, have become a pervasive
instrument in financial engineering. Besides the Swaptions described in the
previous section, this class of instruments includes index options and
exchange options in the equity markets, or yield curve options and spread
options in fixed income markets. In these markets, baskets provide raw
information about the correlation between instruments which is central to
the pricing of exotic derivatives. In this section, we detail an efficient
pricing approximation technique that leads to very natural closed-form
basket pricing formulas with excellent precision results.

The classical ''noise addition in decibels'' order zero lognormal
approximation was studied by \cite{Huyn94}, \cite{Musi97} and
\cite{Brac99} when the underlying instruments follow a
\cite{Blac73} like lognormal diffusion. Here, we approximate the
price of a basket using stochastic expansion techniques similar to
those used by \cite{Four97} or \cite{fouq00} \ on other stochastic
volatility problems. This provides a theoretical justification for
the classical price approximation and allows us to compute
additional terms, better accounting for the stochastic nature of
the basket volatility. In fact, the first correction can be
interpreted as a first order approximation of the hedging tracking
error as defined in \cite{ElKa98} .

\subsection{Generic multivariate lognormal model}

We suppose that the market is composed of $n$ risky assets $%
S_{t}^{i},i=1,\ldots ,n$ plus one riskless asset $M_{t}$. We assume that
these processes are defined on a probability space $(\Omega ,F,\mathbf{Q})$
and are adapted to the natural filtration $\left\{ F_{t},0\leq t\leq
T\right\} $. We suppose that there exists a forward martingale measure $%
\mathbf{Q}$ as defined in \cite{Elka95b} (the notation
$\mathbf{Q}$ is left voluntarily non specific for our purposes
here because it can either be associated with the forward market
of maturity $T$ and constructed by taking the savings account as a
num\'eraire or it could be the level payment induced martingale
measure as in the Swaption pricing formulas treated in the first
section). In this market, the dynamics of the forwards $F_{t}^{i}$
are given by $dF_{s}^{i}=F_{s}^{i}\sigma _{s}^{i}dW_{s}$ and
$M_{s}=1$ for $s\in \lbrack t,T]$, where $W_{t}$ is a
$d$-dimensional $\mathbf{Q}$-Brownian motion adapted to the
filtration $\left\{ F_{t}\right\} $ and $\sigma _{s}=\left( \sigma
_{s}^{i}\right) _{i=1,\ldots ,n}\in $ $\reals^{n\times d}$ is the
volatility matrix and we note $\Gamma _{s}\in \reals^{n\times n}$
the corresponding covariance matrix defined as $\left( \Gamma
_{s}\right) _{i,j}=<\sigma _{s}^{i},\sigma _{s}^{j}>$. We study
the pricing of an option on a basket of forwards given by
$F_{t}^{\omega }=\sum_{i=1}^{n}\omega _{i}F_{t}^{i}$ where $\omega
=\left( \omega _{i}\right) _{i=1,\ldots ,n}\in \reals^{n}$. The
terminal payoff of this option at maturity $T$ is computed as:
\begin{equation*}
h\left( F_{T}^{\omega }\right) =\left( \sum_{i=1}^{n}\omega
_{i}F_{T}^{i}-k\right) ^{+}
\end{equation*}
for a strike price $k$. The key observation at the origin of the following
approximations is that the basket process dynamics are close to lognormal.
The simple formula for basket prices that we will get is specifically
centered around a deterministic approximation of the basket volatility:
\begin{equation}
dF_{s}^{\omega }=F_{s}^{\omega }\left(
\sum_{i=1}^{n}\widehat{\omega } _{i,s}\sigma _{s}^{i}\right)
dW_{s}\text{ \ with \ }\widehat{\omega }_{i,s}=\frac{\omega
_{i}F_{s}^{i}}{\sum_{i=1}^{n}\omega _{i}F_{s}^{i}}
\label{basketdynamics}
\end{equation}

\subsection{Diffusion approximation}

The classical order zero formula approximates the sum of
lognormals as a lognormal variable while matching the two first
moments. This method has its origin in the electrical engineering
literature as a classic problem in signal processing where it
represents, for example, the addition of noise in decibels (see
\cite{schw82} among others). The same method was then used in
finance by \cite{Huyn94}, \cite{Musi97} for equity baskets or
\cite{Brac99} for Swaptions. Here we justify this empirical result
and look for an extra term that better accounts for the (mildly)
stochastic nature of the basket volatility and improves the
pricing approximation outside of the money. The approximation
above simply expresses the fact that if all the forward volatility
vectors were equal then the basket diffusion would then be exactly
lognormal. It is then quite natural to look for an extra term by
developing the above approximation around the central first-order
volatility vector $\sum_{j=1}^{n}\widehat{\omega }_{i,t}\sigma
_{s}^{j}$. As in the previous section, we first define the
residual volatility $\xi _{s}^{i}$ as the difference between the
original volatility $\sigma _{s}^{i}$ and the central basket
volatility $\sum_{j=1}^{n}\widehat{\omega }_{j,t}\sigma _{s}^{j}$
and we set $\xi _{s}^{i}=\sigma
_{s}^{i}-\sum_{j=1}^{n}\widehat{\omega }_{j,t}\sigma _{s}^{j}$,
for $i=1,...,n$ and $s\in \lbrack t,T]$. We also note $\sigma
_{s}^{\omega }=\sum_{j=1}^{n}\widehat{\omega }_{j,t}\sigma
_{s}^{j}$ (notice that $\sigma _{s}^{\omega }$ is
$F_{t}-measurable$).

We can write the dynamics of the basket $F_{s}^{\omega }$ in terms
of $\widehat{\omega }_{i,s}$ and the residual volatilities $\xi
_{s}^{i}$. Remember that for $s\in \lbrack t,T]$ we have
$\widehat{\omega }_{j,s}\geq 0$ with
$\sum_{j=1}^{n}\widehat{\omega }_{j,s}=1$, hence $\sigma
_{s}^{\omega }$ is a convex combination of the $\sigma _{s}^{j}$
and $\sum_{j=1}^{n}\widehat{\omega }_{j,s}\xi _{s}^{j}$ is a
convex combination of the residual volatilities $\xi _{s}^{j}$
with $\sum_{j=1}^{n}\widehat{\omega }_{j,t}\xi _{t}^{j}=0$. As
this term tends to be very small, we will now compute the small
noise expansion of the basket Call price around such small values
of $\sum_{j=1}^{n}\widehat{\omega }_{j,s}\xi _{s}^{j}.$ We first
write
\begin{equation}
\left\{
\begin{array}{l}
dF_{s}^{\omega ,\varepsilon }=F_{s}^{\omega ,\varepsilon }\left( \sigma
_{s}^{\omega }+\varepsilon \sum_{j=1}^{n}\widehat{\omega }_{j,s}\xi
_{s}^{j}\right) dW_{s} \\
\\
d\widehat{\omega }_{i,s}^{\varepsilon }=\widehat{\omega }_{i,s}^{\varepsilon
}\left( \xi _{s}^{i}-\varepsilon \sum_{j=1}^{n}\widehat{\omega }_{j,s}\xi
_{s}^{j}\right) \left( dW_{s}+\sigma _{s}^{\omega }ds+\varepsilon
\sum_{j=1}^{n}\widehat{\omega }_{j,s}\xi _{s}^{j}ds\right)
\end{array}
\right.  \label{JointDyn}
\end{equation}
and develop around small values of $\varepsilon >0.$ As in \cite{Four97}, we
want to evaluate the price and develop its series expansion in $\varepsilon $
around 0.
\begin{equation*}
C^{\varepsilon }=E\left[ \left( F_{T}^{\omega ,\varepsilon }-k\right)
^{+}|\left( F_{t}^{\omega },\widehat{\omega }_{t}\right) \right] \text{ \ \
with \ }C^{\varepsilon }=C^{0}+C^{(1)}\varepsilon +C^{(2)}\frac{\varepsilon
^{2}}{2}+o(\varepsilon ^{2})
\end{equation*}
We can now get the order zero term as the classical basket approximation,
which corresponds to that in \cite{Huyn94}, \cite{Musi97} or \cite{Brac00}.

\begin{proposition}
The first term $C^{0}$ is given by the \cite{Blac73} formula. In this simple
approximation, the basket call price is given by:
\begin{equation}
C^{0}=BS(T,F_{t}^{\omega },V_{t,T})=F_{t}^{\omega
}N(h(V_{t,T}))-\kappa N\left( h(V_{t,T})-\sqrt{V_{t,T}}\right)
\label{vzero}
\end{equation}
where
\begin{equation*}
h\left( V_{t,T}\right) =\frac{\left( \ln \left(
\frac{F_{t}^{\omega }}{\kappa }\right) +\frac{1}{2}V_{t,T}\right)
}{\sqrt{V_{t,T}}}\text{ \ \ and \ \ } V_{t,T}=\int_{t}^{T}\left\|
\sigma _{s}^{\omega }\right\| ^{2}ds
\end{equation*}
where the variance can also be computed as
$V_{t,T}=\int_{t}^{T}Tr\left( \Omega _{t}\Gamma _{s}\right) ds$
with $\Omega _{t}=\widehat{\omega }_{t}\widehat{\omega }_{t}^{T}$.
\label{prop-c-zero}
\end{proposition}

\begin{proof}
Because for $s\in \lbrack t,T]$ we have $\widehat{\omega }_{j,s}\geq 0$ with
$\sum_{j=1}^{n}\widehat{\omega }_{j,s}=1$, as in \cite{Four97} or \cite
{fouq00}\ we can compute $C^{0}$ by solving the limit P.D.E.:
\begin{equation*}
\left\{
\begin{array}{l}
\frac{\partial C^{0}}{\partial s}+\left\| \sigma _{s}^{\omega
}\right\| ^{2}\frac{x^{2}}{2}\frac{\partial ^{2}C^{0}}{\partial x^{2}}=0 \\
\\
C^{0}=\left( x-K\right) ^{+}\text{ for }s=T
\end{array}
\right.
\end{equation*}
hence the above result. Finally $Tr\left( \Omega _{t}\Gamma
_{s}\right) =\sum_{i=1}^{n}\sum_{j=1}^{n}\widehat{\omega
}_{i,t}\widehat{\omega }_{j,t}<\sigma _{s}^{j},\sigma
_{s}^{i}>=\left\| \sigma _{s}^{\omega }\right\| ^{2}$ allows us to
rewrite the variance as the inner product of $\Omega _{t}$ and
$\Gamma _{s}$.
\end{proof}

We have recovered the classical order zero approximation, we can now look
for an extra term by solving for $C^{(1)}$.

\begin{lemma}
Suppose that the underlying dynamics are described by (\ref{JointDyn}). The
first order term $C^{(1)}(s,x,y)$ can be computed by solving:
\begin{eqnarray}
0 &=&\frac{\partial C^{(1)}}{\partial s}+\left\| \sigma
_{s}^{\omega }\right\| ^{2}\frac{x^{2}}{2}\frac{\partial
^{2}C^{(1)}}{\partial x^{2}}+\sum_{j=1}^{n}\left\langle \xi
_{s}^{j},\sigma _{s}^{\omega }\right\rangle xy_{j}\frac{\partial
^{2}C^{(1)}}{\partial x\partial y_{j}}  \label{bigpde}
\\
&&+\sum_{j=1}^{n}\left\| \xi _{s}^{j}\right\|
^{2}\frac{y_{j}^{2}}{2}\frac{\partial ^{2}C^{(1)}}{\partial
y_{j}^{2}}+\sum_{j=1}^{n}\left\langle \xi _{s}^{j},\sigma
_{s}^{\omega }\right\rangle y_{j}\frac{\partial C^{(1)}}{\partial
y_{j}}+\sum_{j=1}^{n}\left\langle \xi _{s}^{j},\sigma _{s}^{\omega
}\right\rangle y_{j}x^{2}\frac{\partial ^{2}C^{0}}{\partial x^{2}}  \notag \\
0 &=&C^{(1)}\text{ for }s=T  \notag
\end{eqnarray}
with $C^{0}=BS(s,x,V_{s})$ given by the \cite{Blac73} formula as in (\ref
{vzero}).
\end{lemma}

\begin{proof}
Let us first detail explicitly the P.D.E. followed by the price
process. With the dynamics given by:
\begin{equation*}
\left\{
\begin{array}{l}
dF_{s}^{\omega ,\varepsilon }=F_{s}^{\omega ,\varepsilon }\left( \sigma
_{s}^{\omega }+\varepsilon \sum_{j=1}^{n}\widehat{\omega }_{j,s}\xi
_{s}^{j}\right) dW_{s} \\
\\
d\widehat{\omega }_{i,s}^{\varepsilon }=\widehat{\omega
}_{i,s}^{\varepsilon }\left( \xi _{s}^{i}-\varepsilon
\sum_{j=1}^{n}\widehat{\omega }_{j,s}^{\varepsilon }\xi
_{s}^{j}\right) \left( dW_{s}+\sigma _{s}^{\omega }ds+\varepsilon
\sum_{j=1}^{n}\widehat{\omega }_{j,s}\xi _{s}^{j}ds\right)
\end{array}
\right.
\end{equation*}
as in \cite{Kara91} we get for
\begin{equation*}
C^{\varepsilon }=E\left[ \left( F_{T}^{\omega ,\varepsilon }-k\right)
^{+}|\left( F_{t}^{\omega },\widehat{\omega }_{t}\right) \right]
\end{equation*}
the corresponding P.D.E. :
\begin{equation*}
\left\{
\begin{array}{l}
L_{0}^{\varepsilon }C^{\varepsilon }=0 \\
C^{\varepsilon }=\left( x-k\right) ^{+}\text{ for }s=T
\end{array}
\right.
\end{equation*}
where $L_{0}^{\varepsilon }$ is given by (with $x$ and $y_{i}$ associated to
$F_{s}^{\omega ,\varepsilon }$ and $\widehat{\omega }_{i,s}$ respectively):
\begin{align*}
L_{0}^{\varepsilon }& =\frac{\partial C^{\varepsilon }}{\partial s}+\left\|
\sigma _{s}^{\omega }+\varepsilon \sum_{j=1}^{n}y_{j}\xi _{s}^{j}\right\|
^{2}\frac{x^{2}}{2}\frac{\partial ^{2}C^{\varepsilon }}{\partial x^{2}} \\
& +\sum_{j=1}^{n}\left( \left\langle \xi _{s}^{j},\sigma
_{s}^{\omega }\right\rangle +\varepsilon
\sum_{k=1}^{n}y_{k}\left\langle \xi _{s}^{j}-\sigma _{s}^{\omega
},\xi _{s}^{k}\right\rangle -\varepsilon ^{2}\left\|
\sum_{k=1}^{n}y_{k}\xi _{s}^{k}\right\| ^{2}\right)
xy_{j}\frac{\partial ^{2}C^{\varepsilon }}{\partial x\partial y_{j}} \\
& +\sum_{j=1}^{n}\left\| \xi _{s}^{j}-\varepsilon
\sum_{k=1}^{n}y_{k}\xi _{s}^{k}\right\|
^{2}\frac{y_{j}^{2}}{2}\frac{\partial ^{2}C^{\varepsilon }}{\partial y_{j}^{2}} \\
& +\sum_{j=1}^{n}\left( \left\langle \xi _{s}^{j},\sigma
_{s}^{\omega }\right\rangle +\varepsilon
\sum_{k=1}^{n}y_{k}\left\langle \xi _{s}^{j}-\sigma _{s}^{\omega
},\xi _{s}^{k}\right\rangle -\varepsilon ^{2}\left\|
\sum_{k=1}^{n}y_{k}\xi _{s}^{k}\right\| ^{2}\right)
y_{j}\frac{\partial C^{\varepsilon }}{\partial y_{j}}
\end{align*}
as in \cite{Four97} we can differentiate this P.D.E. with respect
to $\varepsilon $ to get:
\begin{eqnarray*}
0 &=&L_{0}^{\varepsilon }C^{(1),\varepsilon }+\left(
2\sum_{j=1}^{n}y_{j}\left\langle \xi _{s}^{j},\sigma _{s}^{\omega
}\right\rangle +2\varepsilon \left\| \sum_{k=1}^{n}y_{k}\xi _{s}^{k}\right\|
^{2}\right) \frac{x^{2}}{2}\frac{\partial ^{2}C^{\varepsilon }}{\partial
x^{2}} \\
&&+\sum_{j=1}^{n}\left( \sum_{k=1}^{n}\left\langle \xi _{s}^{j}-\sigma
_{s}^{\omega },\xi _{s}^{j}\right\rangle -2\varepsilon \left\|
\sum_{k=1}^{n}y_{k}\xi _{s}^{k}\right\| ^{2}\right) xy_{j}\frac{\partial
^{2}C^{\varepsilon }}{\partial x\partial y_{j}} \\
&&+\sum_{j=1}^{n}\left( -2\sum_{k=1}^{n}y_{k}\left\langle \xi _{s}^{j},\xi
_{s}^{k}\right\rangle +2\varepsilon \left\| \sum_{k=1}^{n}y_{k}\xi
_{s}^{k}\right\| ^{2}\right) \frac{y_{j}^{2}}{2}\frac{\partial
^{2}C^{\varepsilon }}{\partial y_{j}^{2}} \\
&&+\sum_{j=1}^{n}\left( \sum_{k=1}^{n}y_{k}\left\langle \xi _{s}^{j}-\sigma
_{s}^{\omega },\xi _{s}^{k}\right\rangle -2\varepsilon \left\|
\sum_{k=1}^{n}y_{k}\xi _{s}^{k}\right\| ^{2}\right) y_{j}\frac{\partial
C^{\varepsilon }}{\partial y_{j}} \\
0 &=&C^{(1),\varepsilon }\text{ for }s=T
\end{eqnarray*}
and again as in \cite{Four97} or \cite{fouq00} we take the limit
as $\varepsilon \rightarrow \infty $ and compute $C^{(1)}$ as the
solution to:
\begin{equation*}
\left\{
\begin{array}{l}
L_{0}^{0}C^{(1)}+\left( \sum_{j=1}^{n}y_{j}\left\langle \xi _{s}^{j},\sigma
_{s}^{\omega }\right\rangle \right) x^{2}\frac{\partial ^{2}C^{0}}{\partial
x^{2}}=0 \\
C^{\varepsilon }=0\text{ for }s=T
\end{array}
\right.
\end{equation*}
which is again, with $C^{0}=BS(T,F_{t}^{\omega },V_{t,T})$ given
by (\ref {vzero}):
\begin{eqnarray*}
0 &=&\frac{\partial C^{(1)}}{\partial s}+\left\| \sigma
_{s}^{\omega }\right\| ^{2}\frac{x^{2}}{2}\frac{\partial
^{2}C^{(1)}}{\partial x^{2}}+\sum_{j=1}^{n}\left\langle \xi
_{s}^{j},\sigma _{s}^{\omega }\right\rangle
xy_{j}\frac{\partial ^{2}C^{(1)}}{\partial x\partial y_{j}} \\
&&+\sum_{j=1}^{n}\left\| \xi _{s}^{j}\right\|
^{2}\frac{y_{j}^{2}}{2}\frac{\partial ^{2}C^{(1)}}{\partial
y_{j}^{2}}+\sum_{j=1}^{n}\left\langle \xi _{s}^{j},\sigma
_{s}^{\omega }\right\rangle y_{j}\frac{\partial C^{(1)}}{\partial
y_{j}}+\sum_{j=1}^{n}\left\langle \xi _{s}^{j},\sigma _{s}^{\omega
}\right\rangle y_{j}x^{2}\frac{\partial ^{2}C^{0}}{\partial x^{2}} \\
0 &=&C^{(1)}\text{ for }s=T
\end{eqnarray*}
which is the desired result.
\end{proof}

We can now compute a closed-form solution to the equation verified
by $C^{(1)}$ using its Feynman-Kac representation.

\begin{proposition}
Suppose that the underlying dynamics are described by (\ref{JointDyn}).

The derivative $C^{(1)}\left( t,F_{t}^{\omega },\left(
\widehat{\omega }_{j,t}\right) _{j=1,...,n}\right) $ can be
computed as:
\begin{eqnarray}
C^{(1)} &=&F_{t}^{\omega
}\int_{t}^{T}\sum_{j=1}^{n}\widehat{\omega
}_{j,t}\frac{\left\langle \xi _{s}^{j},\sigma _{s}^{\omega
}\right\rangle }{\sqrt{ V_{t,T}}}\exp \left(
2\int_{t}^{s}\left\langle \xi _{u}^{j},\sigma
_{u}^{\omega }\right\rangle du\right)  \label{firstorderterm} \\
&&n\left( \frac{\ln \frac{F_{t}^{\omega
}}{K}+\int_{t}^{s}\left\langle \xi _{u}^{j},\sigma _{u}^{\omega
}\right\rangle du+\frac{1}{2}V_{t,T}}{\sqrt{V_{t,T}}}\right) ds
\notag
\end{eqnarray}
\label{prop-c-one}
\end{proposition}

\begin{proof}
The limiting diffusions are given by:
\begin{eqnarray*}
F_{s}^{\omega ,0} &=&F_{t}^{\omega }\exp \left( \int_{t}^{s}\sigma
_{u}^{\omega }dW_{u}-\frac{1}{2}\int_{t}^{s}\left\| \sigma _{u}^{\omega
}\right\| ^{2}du\right) \\
\widehat{\omega }_{j,s}^{0} &=&\widehat{\omega }_{j,t}\exp \left(
\int_{t}^{s}\xi_{u}^{j}dW_{u}+\int_{t}^{s}\left( \left\langle \xi
_{u}^{j},\sigma _{u}^{\omega }\right\rangle -\frac{1}{2}\left\|
\xi_{u}^{j}\right\| ^{2}\right) du\right)
\end{eqnarray*}
and because $C^{(1)}$ solves the P.D.E. (\ref{bigpde}) in the above lemma,
with
\begin{equation*}
\frac{\partial ^{2}C_{s}^{0}}{\partial
x^{2}}=\frac{n(h(x,V_{s,T}))}{x\sqrt{V_{s,T}}}\text{ \ \ where \ \
}n(x)=\frac{1}{\sqrt{2\pi }}\exp \left( -\frac{1}{2}x^{2}\right)
\end{equation*}
We can write the Feynman-Kac representation of the solution to
(\ref{bigpde}) with terminal condition zero as:
\begin{equation*}
C^{(1)}=\int_{t}^{T}E\left[ \sum_{j=1}^{n}\left\langle \xi
_{s}^{j},\sigma _{s}^{\omega }\right\rangle \widehat{\omega
}_{j,s}^{0}F_{s}^{\omega ,0}\frac{n(h(V_{s,T},F_{s}^{\omega
,0}))}{\sqrt{V_{s,T}}}\right] ds
\end{equation*}
where
\begin{equation*}
h\left( u,v\right) =\frac{\left( \ln \left( \frac{v}{\kappa
}\right) +\frac{1}{2}u\right) }{\sqrt{u}}\text{ \ \ with \
}V_{s,T}=\int_{s}^{T}\left\| \sigma _{u}^{\omega }\right\| ^{2}du
\end{equation*}
Hence we can directly compute $C^{(1)}$ as:
\begin{align*}
& C^{(1)}=F_{t}^{\omega }\int_{t}^{T}\sum_{j=1}^{n}\widehat{\omega
}_{j,t}\left\langle \xi _{s}^{j},\sigma _{s}^{\omega
}\right\rangle \exp \left( \int_{t}^{s}-\frac{1}{2}\left\| \xi
_{u}^{j}-\sigma _{u}^{\omega
}\right\| ^{2}du\right) \\
& E\left[ \frac{\exp \left( \int_{t}^{s}\left( \sigma _{u}^{\omega
}+\xi _{u}^{j}\right) dW_{u}\right) }{\sqrt{V_{s,T}}}n\left(
\frac{\ln \frac{F_{t}^{\omega }}{K}+\int_{t}^{s}\sigma
_{u}^{\omega
}dW_{u}-\frac{1}{2}V_{t,s}+\frac{1}{2}V_{s,T}}{\sqrt{V_{s,T}}}\right)
\right] ds
\end{align*}
which is, using the Cameron-Martin formula:
\begin{eqnarray*}
&&C^{(1)}=F_{t}^{\omega }\int_{t}^{T}\sum_{j=1}^{n}\widehat{\omega
}_{j,t}\frac{\left\langle \xi _{s}^{j},\sigma _{s}^{\omega
}\right\rangle \exp \left( 2\int_{t}^{s}\left\langle \xi
_{u}^{j},\sigma _{u}^{\omega
}\right\rangle du\right) }{\sqrt{V_{s,T}}} \\
&&E\left[ n\left( \frac{\ln \frac{F_{t}^{\omega
}}{K}+\int_{t}^{s}\sigma _{u}^{\omega
}dW_{u}+\int_{t}^{s}\left\langle \xi _{u}^{j},\sigma _{u}^{\omega
}\right\rangle du+\frac{1}{2}V_{t,T}}{\sqrt{V_{s,T}}}\right)
\right] ds
\end{eqnarray*}
and because for $g=N(a,b^{2}):$
\begin{equation*}
E[n(g)]=\frac{1}{\sqrt{b^{2}+1}}n\left( \frac{a}{\sqrt{b^{2}+1}}\right)
\end{equation*}
we get:
\begin{eqnarray*}
C^{(1)} &=&F_{t}^{\omega
}\int_{t}^{T}\sum_{j=1}^{n}\widehat{\omega
}_{j,t}\frac{\left\langle \xi _{s}^{j},\sigma _{s}^{\omega
}\right\rangle }{\sqrt{\left( V_{t,s}+V_{s,T}\right) }}\exp \left(
2\int_{t}^{s}\left\langle \xi
_{u}^{j},\sigma _{u}^{\omega }\right\rangle du\right) \\
&&n\left( \frac{\ln \frac{F_{t}^{\omega
}}{K}+\int_{t}^{s}\left\langle \xi _{u}^{j},\sigma _{u}^{\omega
}\right\rangle du+\frac{1}{2}V_{t,T}}{\sqrt{\left(
V_{t,s}+V_{s,T}\right) }}\right) ds
\end{eqnarray*}
which is the desired result.
\end{proof}

It is possible to compute the order two term explicitly but the computations
are a bit longer. We will show below that this result can be interpreted as
a correction accounting for the misspecification of the volatility induced.

\subsection{Robustness interpretation}

The basket dynamics are essentially that of an almost lognormal
process with a mildly stochastic volatility. By approximating
these dynamics with a true lognormal process, we will make a small
''tracking error'' in hedging by computing the delta using an
incorrect specification of the volatility. As in \cite{ElKa98}, we
can compute this tracking error almost explicitly. Suppose that
$\Pi _{\sigma _{s}^{\omega },s}$ is the value at time $s$ of a
self-financing delta hedging portfolio computed using the
approximate volatility $\sigma _{s}^{\omega }$. As the volatility
in this delta computation is only approximately equal to the
volatility driving the underlying assets, there will be a small
hedging tracking error $e_{s}$ computed as $e_{s}=P_{\sigma
_{s}^{\omega },s}-\Pi _{\sigma _{s}^{\omega },s}$ for $s\in
\lbrack t,T]$, where $P_{\sigma _{s}^{\omega },s}$ is the price of
the option at time $s$, computed using the approximate volatility
$\sigma _{s}^{\omega }$. Of course, we know that $P_{\sigma
_{s}^{\omega },T}=\left( F_{t}^{\omega }-K\right) ^{+}$ and we can
understand $E[e_{s}]$ as a price correction accounting for the
volatility misspecification. From \cite{ElKa98} we know that we
can compute this (exact) tracking error explicitly as:
\begin{equation}
e_{T}=\frac{1}{2}\int_{t}^{T}\left( \left\|
\sum_{i=1}^{n}\widehat{\omega }_{i,s}\sigma _{s}^{i}\right\|
^{2}-\left\| \sigma _{s}^{\omega }\right\| ^{2}\right) \left(
F_{s}^{\omega }\right) ^{2}\frac{\partial ^{2}C^{0}(F_{s}^{\omega
},V_{t,T})}{\partial x^{2}}ds  \label{track-error}
\end{equation}
From the computation of $C^{(1)}$ in the previous part we know:
\begin{equation*}
C^{(1)}=\int_{t}^{T}E\left[ \sum_{j=1}^{n}\left\langle \xi
_{s}^{j},\sigma _{s}^{\omega }\right\rangle \widehat{\omega
}_{j,s}F_{s}^{\omega }\frac{n(h(V_{s,T},F_{s}^{\omega
}))}{\sqrt{V_{s,T}}}\right] ds
\end{equation*}
With $\sigma _{s}^{i}=\sigma _{s}^{\omega }+\xi _{s}^{i}$, and
because $\sum_{i=1}^{n}\widehat{\omega }_{i,s}=1,$ we rewrite
(\ref{track-error}) as:
\begin{eqnarray*}
e_{T} &=&\int_{t}^{T}\left( \left\langle
\sum_{i=1}^{n}\widehat{\omega }_{i,s}\sigma _{s}^{i}-\sigma
_{s}^{\omega },\sigma _{s}^{\omega }\right\rangle \right) \left(
F_{s}^{\omega }\right) ^{2}\frac{\partial
^{2}C^{0}(F_{s}^{\omega },V_{t,T})}{\partial x^{2}}ds \\
&&+\frac{1}{2}\int_{t}^{T}\left( \left\|
\sum_{i=1}^{n}\widehat{\omega }_{i,s}\sigma _{s}^{i}-\sigma
_{s}^{\omega }\right\| ^{2}\right) \left( F_{s}^{\omega }\right)
^{2}\frac{\partial ^{2}C^{0}(F_{s}^{\omega },V_{t,T})}{\partial
x^{2}}ds
\end{eqnarray*}
The first order expansion of $e_{T}$ for small values of $\xi _{s}^{i}$
gives:
\begin{equation*}
e_{T}^{(1)}=\int_{t}^{T}\left\langle \sum_{i=1}^{n}\widehat{\omega
}_{i,s}\xi _{s}^{i},\sigma _{s}^{\omega }\right\rangle \left(
F_{s}^{\omega }\right) ^{2}\frac{\partial ^{2}C^{0}(F_{s}^{\omega
},V_{t,T})}{\partial x^{2}}ds
\end{equation*}
writing the value of the Gamma explicitly, we get:
\begin{equation*}
e_{T}^{(1)}=\int_{t}^{T}\sum_{i=1}^{n}\left\langle \xi
_{s}^{i},\sigma _{s}^{\omega }\right\rangle \widehat{\omega
}_{i,s}F_{s}^{\omega }\frac{n(h(V_{s,T},F_{s}^{\omega
}))}{\sqrt{V_{s,T}}}ds
\end{equation*}
and finally $C^{(1)}=E\left[ e_{T}^{(1)}\right] $. This means that the first
order correction in the basket price approximation can also be interpreted
as the expected value of\ the first order tracking error approximation for
small values of the residual volatility $\xi _{s}^{i}$. This validates the
price approximation in terms of \textit{both pricing and hedging performance}%
. To make the link with section two explicit, we now write the order zero
approximation in the particular case of Swaption pricing.

\subsection{Swaption price approximation}

If we go back to the particular Swaption pricing problem developed in
section two, the result above allows us to approximate the price of a
Swaption.

\begin{proposition}
Using the above approximations, the price of a payer Swaption with
maturity $T$ and strike $\kappa $, written on a Forward Swap
starting at $T$ with maturity $T_{N}$ is given at time $t\leq T$ \
by the Black formula plus a correction term:
\begin{equation}
Swaption_{t}=Level(t)\left( swap(t)N(h)-\kappa
N(h-\sqrt{V_{t,T}})\right) +Level(t)C^{(1)}
\end{equation}
with
\begin{equation*}
h=\frac{\left( \ln \left( \frac{swap(t)}{\kappa }\right)
+\frac{1}{2}V_{t,T}\right) }{\sqrt{V_{t,T}}}\text{ \ \ with \ \
}V_{t,T}=\int_{t}^{T}\left\| \gamma ^{\omega }(s)\right\| ^{2}ds
\end{equation*}
where $swap(t,T,T_{N})$ is the market value of the Forward swap today with
\begin{equation*}
\hat{\omega}_{i}(t)=\omega _{i}(t)\frac{K(t,T_{i})}{swap(t)}\text{
\ \ and \ }\gamma
^{\omega}(s)=\sum_{i=1}^{N}\hat{\omega}_{i}(t)\gamma (s,T_{i}-s)
\end{equation*}
and
\begin{eqnarray*}
C^{(1)} &=&\int_{t}^{T}\sum_{j=1}^{n}\widehat{\omega
}_{j}(t)\frac{\left\langle \xi (s,T_{j}-s),\gamma ^{\omega
}(s)\right\rangle }{\sqrt{V_{t,T}}}\exp \left(
2\int_{t}^{s}\left\langle \xi (s,T_{j}-s),\gamma
^{\omega }(s)\right\rangle du\right) \\
&&n\left( \frac{\ln \frac{Level(t)}{K}+\int_{t}^{s}\left\langle
\xi (s,T_{j}-s),\gamma ^{\omega }(s)\right\rangle
du+\frac{1}{2}V_{t,T}}{\sqrt{V_{t,T}}}\right) ds
\end{eqnarray*}
where $\xi (s,T_{i}-s)=\gamma (s,T_{i}-s)-\gamma ^{\omega }(s)$.
\end{proposition}
\begin{proof}
This is a direct consequence of proposition (\ref{prop-c-zero})
and (\ref{prop-c-one}), with $F_{s}^{i}=K^S(s,T_i)$ and $\sigma
_{s}^{i}=\gamma(s,T_{i}-s)$.
\end{proof}

In the last section, we will study the practical precision of this
approximation by comparing the price obtained using the formulas above with
the price obtained by Monte-Carlo simulations in both the Libor Market model
and in the generic multidimensional \cite{Blac73} model.

\section{Libor market model calibration}

In this section, we detail the calibration problem and its resolution by
semidefinite programming techniques. For a general overview of semidefinite
programming algorithms see \cite{Nest94} or \cite{Vand96}. Because it
provides sufficient precision in most market conditions, we will use the
order zero approximation here (if the rates become less correlated and the
relative variance of the second factor increases, we can always replace $%
\Omega _{t}$ below by a new matrix, factoring in the first order price
correction). Let us write the market variance in the approximation obtained
in the last section as\ a function of the scalar product of the forward
rates covariance matrix and a matrix computed from market data on the Swap
weights:
\begin{align}
V_{t,T}& =\int_{t}^{T}\left\|
\sum_{i=1}^{N}\hat{\omega}_{i}(t)\gamma (s,T_{i}-s)\right\|
^{2}ds=\int_{t}^{T}\left(
\sum_{i=1}^{N}\sum_{j=1}^{N}\hat{\omega}_{i}(t)\hat{\omega}_{j}(t)\left\langle
\gamma (s,T_{i}-s),\gamma
(s,T_{j}-s)\right\rangle \right) ds  \notag \\
& =\int_{t}^{T}Tr\left( \Omega _{t}X_{s}\right) ds
\end{align}
where $\Omega _{t},X_{t}\in \reals^{N\times N},t\in \lbrack 0,T]$ are
positive semidefinite symmetric matrixes defined by:
\begin{equation*}
\Omega _{t}=\hat{\omega}(t)\hat{\omega}(t)^{T}=\left(
\hat{\omega}_{i}(t)\hat{\omega}_{j}(t)\right) _{i,j\in \lbrack
1,N]}\succeq 0\text{ \ \ and \ \ }X_{s}=\left( \left\langle \gamma
(s,T_{i}-s),\gamma (s,T_{j}-s)\right\rangle \right) _{i,j\in
\lbrack 1,N]}\succeq 0
\end{equation*}
i.e. $X_{s}$ is the covariance matrix of the forward rates (Gram matrix of
the $\gamma (s,T_{i}-s)$ vectors). This shows that the cumulative market
variance of a particular Swaption can be written as a \textit{linear
functional} of the Forward rates covariance matrix. With $\sigma
_{market,k}^{2}T_{k}$ for $k=1,...,M$, the market cumulative variance for
the Swaption of maturity $T_{k}$ as inputs, the calibration problem can then
be written as an infinite-dimensional \textit{linear matrix inequality}
(L.M.I.) :
\begin{equation}
\begin{tabular}{ll}
$\text{Find}$ & $X_{s}$ \\
$\text{s.t.}$ & $Tr\left( \Omega _{t}\left( \int_{t}^{T}X_{s}ds\right)
\right) =\sigma _{market,k}^{2}T_{k}\text{ \ for }k=(1,...,M)$ \\
& $X_{s}\succeq 0$%
\end{tabular}
\label{calibLMI}
\end{equation}
in the variable $X_{s}:\reals_{+}\rightarrow \symm_{n}$, where the
matrix $\Omega _{t}$ is quoted by the market today.

Because the market variance constraints are linear with respect to
the underlying variable $X_{s}$ and the set of positive
semidefinite matrixes is a convex cone, we find that the general
calibration problem is convex and given a convex objective
function, it has a unique global solution. For simplicity now and
to keep the focus on the problem geometry, we discretize $X_{s}$
with a $\delta $ frequency and make the common (but not necessary
here) simplifying assumption that although the forward rates
volatilities are not stationary, their instantaneous correlation
is, hence the volatility function take a quasi-stationary form
$\gamma (s,x)=\sigma (s)\eta (x)$ with $\sigma $ and $\eta $ such
that $\sigma (s)=\sigma (\frac{1}{\delta }\left\lfloor \delta
s\right\rfloor ),$ $\eta (u)=\eta (\frac{1}{\delta }\left\lfloor
\delta u\right\rfloor )$ and $\sigma (s)=\eta (s)=0$ when $s\leq
0$. The expression of the market cumulative variance then becomes
$V_{t,T}=\sum_{i=t}^{T}\delta Tr\left( \Omega _{t}X_{i}\right)$.
We can account for Bid-Ask spreads in the market data by relaxing
the constraints as:
\begin{equation}
\begin{tabular}{ll}
$\text{Find}$ & $X_{i}$ \\
$\text{s}$.t. & $\sigma _{Bid,k}^{2}T_{k}\text{ }\leq
\sum_{i=t}^{T}\delta Tr\left( \Omega _{t,k}X_{i}\right) \leq
\sigma _{Ask,k}^{2}T_{k}\text{ for }
k=1,...,M$ \\
& $X_{i}\succeq 0\text{ for }i=0,...,T$
\end{tabular}
\label{DiscCalib}
\end{equation}
where we have set $X_{i}=\left( \sigma ^{2}(s)\left\langle \eta
(T_{i}-s),\eta (T_{j}-s)\right\rangle \right) _{i,j\in \lbrack 1,N]}\succeq
0 $ (keeping in mind that the vectors $\eta (T_{i}-s)$ creating this matrix
''shift'' from period to period). Numerical packages such as SEDUMI by \cite
{Stur99} (for symmetric cone programming) solve these problems with
excellent complexity bounds similar to those obtained for linear programs
(see \cite{Nest98}).

\subsection{Applications}

In general, the calibration problem gives an entire set of solutions.
Different choices of convex objectives are detailed below.

\subsubsection{Bounds on other Swaptions}

One of the most simple choices of objective matrix $C$ is to set it to
another Swaptions associated matrix $\Omega _{T_{i}}$. The calibration
problem finds the parameters for the Libor market model that gives either a
minimum or a maximum arbitrage-free price (within the BGM framework) to the
considered Swaption while matching a certain set of market prices on other
Caps and Swaptions (see \cite{dasp02a} and \cite{dasp02b}).

\subsubsection{Distance to a target covariance matrix}

Let $A$ be a target covariance matrix (for example, a previous calibration
result or an historical estimate), we can minimize $\left\| A-X\right\| $
under the constraints in (\ref{DiscCalib}). If $\left\| .\right\| $ is the
spectral or Euclidean norm, this is a symmetric cone program and can be
solved as in \cite{Nest98} or \cite{Stur99}.

\subsubsection{Maximum entropy}

In the spirit of \cite{Avel87}, let $P$ be a covariance matrix representing
prior information on the distribution of Forwards, as in \cite{vand98} we
can minimize $-\ln \det \left( X\right) +Tr(P^{-1}X)$ to find the maximum
relative entropy solution to the calibration problem.

\subsubsection{Smoothness constraints}

It is sometimes desirable to impose smoothness objectives on the calibration
problem to reflect the fact that market operators will tend to price
similarly the variance of two products with close characteristics. A common
way of smoothing the solution is to minimize the surface of the covariance
matrix that we approximate here by:
\begin{equation*}
S=\sum_{i,j\in \lbrack 2,n]}\left\| \Delta _{i,j}X\right\| ^{2}\text{ \ \
where \ \ }\Delta _{i,j}X=\left(
\begin{array}{c}
X_{i,j}-X_{i-1,j} \\
X_{i,j}-X_{i,j-1}
\end{array}
\right)
\end{equation*}
Again, this is a symmetric cone program.

\subsubsection{Calibration stabilization: a Tikhonov regularization}

Along the lines of \cite{Cont01}, we can explore the impact of the
smoothness constraints introduced above. We can think of the calibration as
an ill-posed inverse problem and write the smooth calibration program as a
\cite{tikh63} regularization of the original problem. If we set, $\sigma
_{Mid,k}^{2}=\left( \sigma _{Bid,k}^{2}+\sigma _{Aks,k}^{2}\right) /2$,
minimizing $\sum_{k=1}^{M}\left\| Tr(\Omega _{k}X)-\left( \sigma
_{Mid,k}^{2}T_{k}\right) \right\| ^{2}+\alpha \left\| X\right\| ^{2}$ will
then directly improve the stability of the calibration problem.

\subsection{Rank Minimization}

Because the calibrated model will be used to compute prices of other
derivatives using mostly Monte-Carlo techniques or trees, it is highly
desirable to get a low rank solution. In general, the matrix solution to the
calibration problem will lie on the border of the semidefinite cone and
hence will be singular but there is no guarantee that the rank will remain
below a certain level. In general (cf.\cite{Vand96}), this problem is
NP-Hard. However, some very efficient heuristical methods (see \cite{Boyd00}
on trace minimization) can produce results with very rapidly decreasing
eigenvalues. In practice and in accordance with prior empirical studies (see
\cite{BGM97}), all solutions (even those with a high rank) tend to have only
one or two dominant eigenvalues with the rest of the spectrum several orders
of magnitude smaller.

\section{Numerical examples}

\subsection{Approximation precision}

\begin{figure}[h]
\begin{center}
\includegraphics[width=.95%
\textwidth]{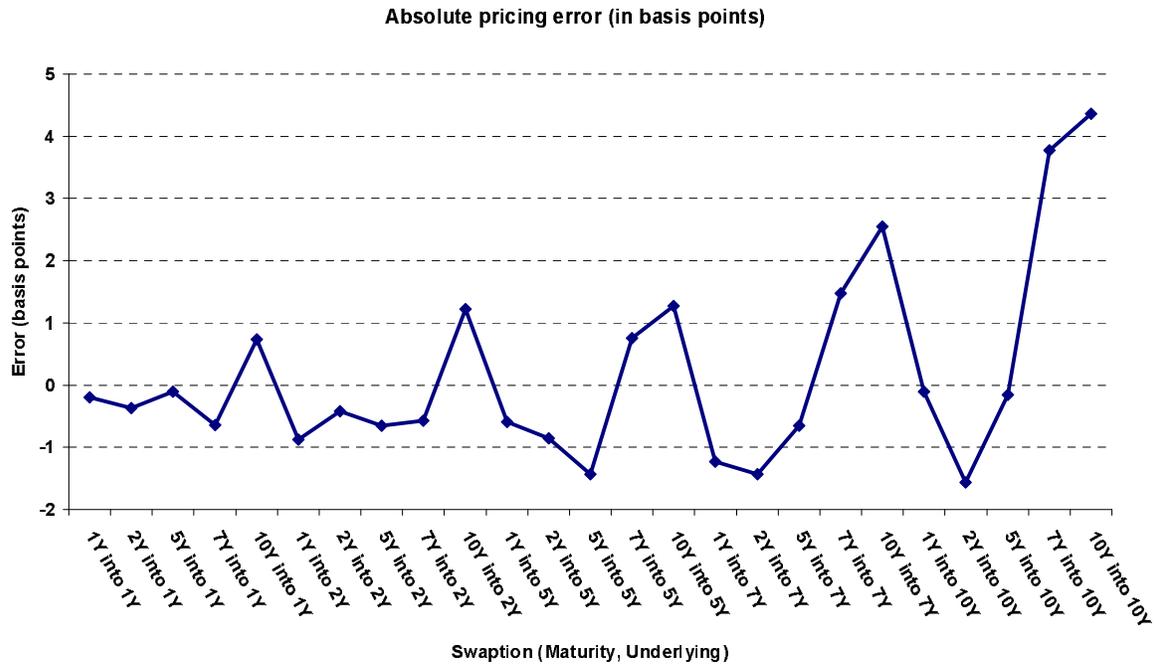}
\end{center}
\caption{Absolute error in the order zero price approximation versus the
Libor market model prices estimated using Monte-Carlo simulation, for
various ATM Swaptions.}
\label{figerrorsignBGM}
\end{figure}

To assess the practical performance of the lognormal swap rate approximation
in the pricing of Swaptions, we will compare the prices obtained for a large
set of key liquid Swaptions using Monte-Carlo simulation and the lognormal
forward swap approximation. We have used the classic Euler discretization
scheme as detailed for example in \cite{Side98}. In figure (\ref
{figerrorsignBGM}), we present \ a plot of the difference between two
distinct sets of Swaption prices in the Libor Market Model. One is obtained
by Monte-Carlo simulation using enough steps to make the 95\% confidence
margin of error always less than 1bp. The second set of prices is computed
using the order zero approximation formula above.
We can notice that the absolute error is increasing in the underlying
maturity of the Swaption and that its sign is not constant. This plot is
based on the prices obtained by calibrating the model to EURO Swaption
prices on November 6 2000 (data courtesy of Paribas Capital Markets,
London). We have used all Cap volatilities and the following Swaptions: 2Y
into 5Y, 5Y into 5Y, 5Y into 2Y, 10Y into 5Y, 7Y into 5Y, 10Y into 2Y, 10Y
into 7Y, 2Y into 2Y, 1Y into 9Y (the motivation behind this choice of
Swaptions is liquidity, all Swaptions in the 10Y diagonal or in 2Y, 5Y, 7Y,
10Y are supposed to be more liquid). The absolute error is always less than
4 bp which is significantly lower than the Bid-Ask spreads.

\begin{figure}[p]
\begin{center}
\parbox{2.5in}{\includegraphics[width=0.4
\textwidth]{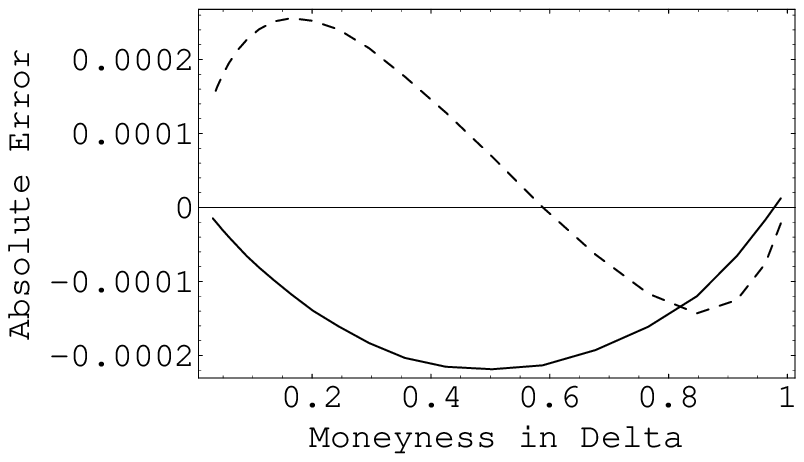}\caption{Order zero (dashed) and
order one (plain) absolute approximation error versus the
multidimensional Black-Scholes basket prices obtained by
simulation for various strikes.}\label{figerrorone5into5}}
\hspace{.25in} \parbox{2.5in}{\includegraphics[width=0.4
\textwidth]{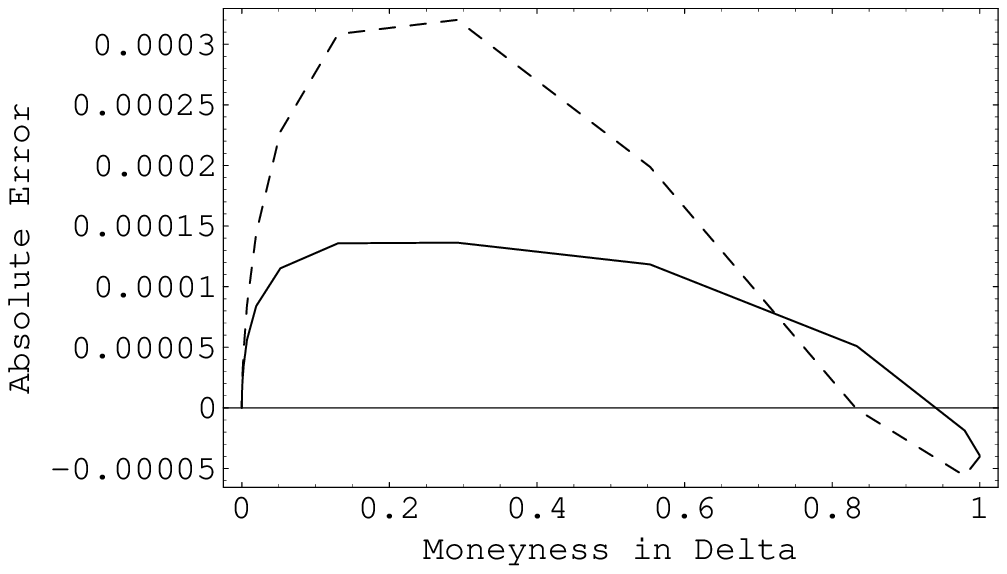}\caption{Order zero (dashed)
and order one (plain) absolute approximation error versus the
multidimensional Black-Scholes basket prices obtained by
simulation for various strikes. (Diagonal covariance
matrix)}\label{figident}}
\end{center}
\end{figure}

In the second figure (\ref{figerrorone5into5}), we plot the error in the
basket pricing formula for a basket of assets, having supposed that the
forwards are all martingale under the same probability measure (hence we
test the precision of the approximations without the error coming from the
forward measures, this is also a test of the formula's precision in an
equity framework). The reference is given by a Monte-Carlo estimate with
40000 steps. The numerical values used here are $F_{0}^{i}=%
\{0.7,0.5,0.4,0.4,0.4\}$, $\omega _{i}=\{0.2,0.2,0.2,0.2,0.2\}$, $T=5$
years, and the covariance matrix is given by:
\begin{equation*}
\frac{11}{100}\left(
\begin{array}{lllll}
0.64 & 0.59 & 0.32 & 0.12 & 0.06 \\
0.59 & 1 & 0.67 & 0.28 & 0.13 \\
0.32 & 0.67 & 0.64 & 0.29 & 0.14 \\
0.12 & 0.28 & 0.29 & 0.36 & 0.11 \\
0.06 & 0.13 & 0.14 & 0.11 & 0.16
\end{array}
\right)
\end{equation*}
The covariance used here comes from an historical estimate and has the
typical level, spread, convexity eigenvector structure. These values are
meant to replicate the pricing of a 5Y into 5Y Swaption without the change
in measure. We can see that the pricing error is less than 2bp with the
order zero approx. and the additional order one term does not provide a
significant benefit. In fact, the order zero term reaches an excellent
precision near the money, a feature that is constantly observed when the
covariance matrix has the structure given above, where the first level
eigenvector accounts for around $90\%$ of the volatility and the model is
close to univariate (as noted in \cite{BGM97}). However, we observe in
figure (\ref{figident}) that the order one approximation does provide a
significant precision improvement when the rates are less correlated.
Finally, in a pure equity case, i.e. when the initial value of the
underlying assets is not significantly smaller than one (an equity basket
option for example), the order one correction very significantly reduces the
relative error, as can be observed in figure (\ref{figequityerrorone}).

\subsection{Calibration}

\begin{figure}[p]
\begin{center}
\parbox{2.5in}{\includegraphics[width=0.42
\textwidth]{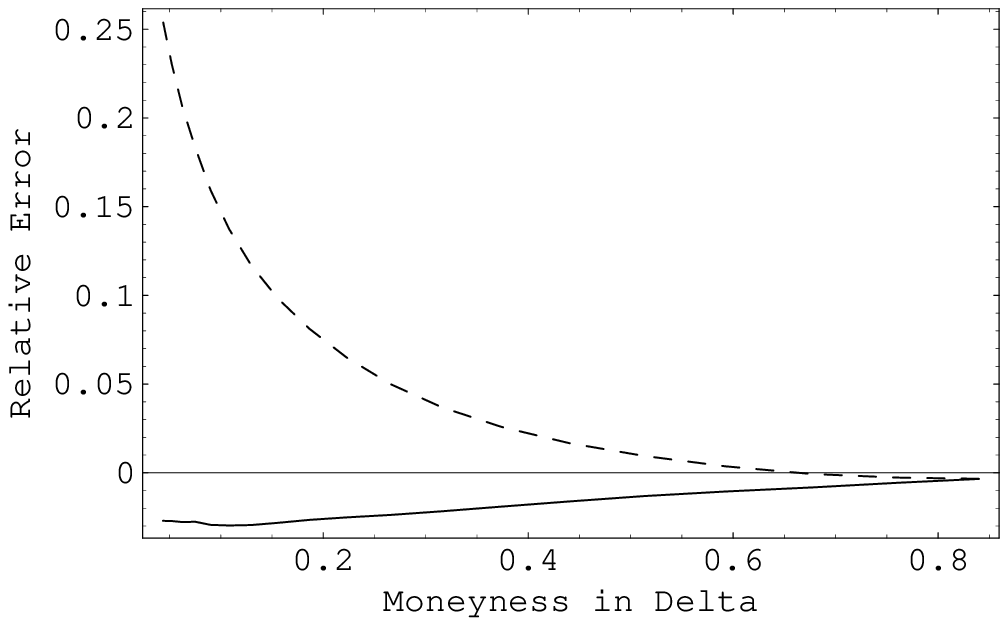}\caption{An equity basket example:
Order zero (dashed) and order one (plain) relative approximation
error versus the multidimensional Black-Scholes basket prices
obtained by simulation for various
strikes.}\label{figequityerrorone}} \hspace{.25in}
\parbox{2.5in}{\includegraphics[width=0.42
\textwidth]{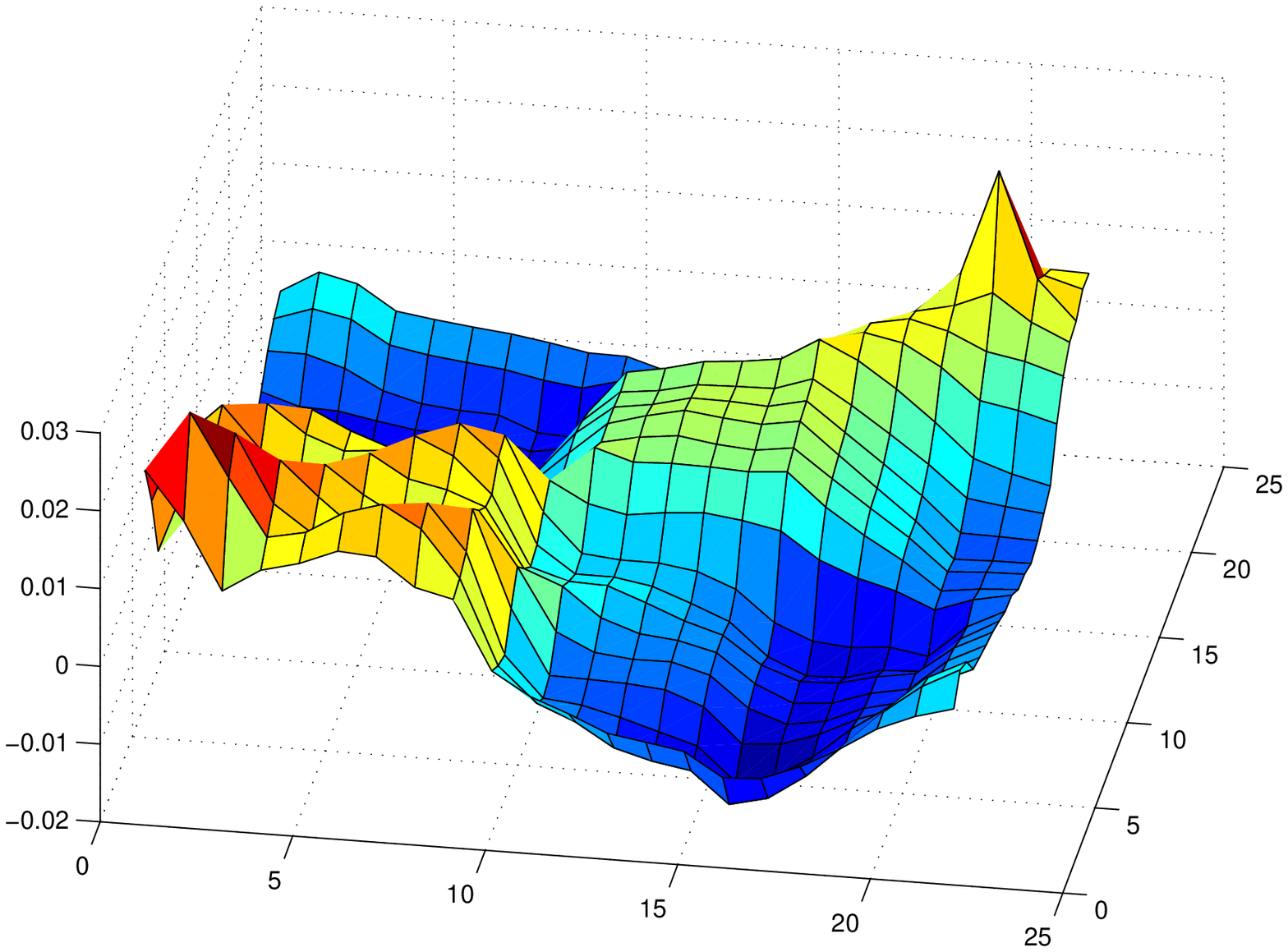}\caption{Smooth calibrated
covariance matrix.}\label{figsmoothmatfactormat}}
\end{center}
\end{figure}

\begin{figure}[p]
\begin{center}
\parbox{2.5in}{\includegraphics[width=0.42
\textwidth]{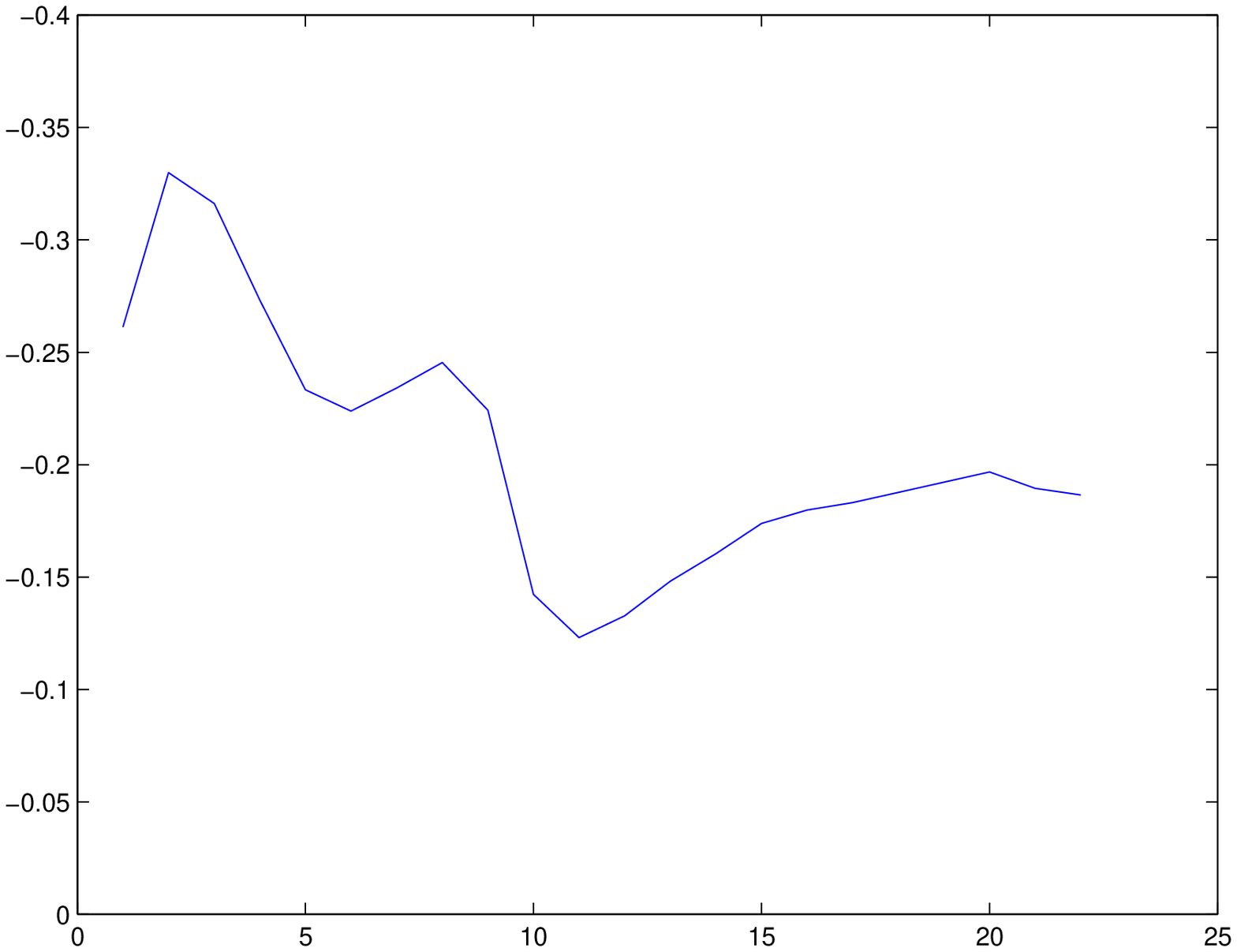} \caption{First eigenvector
''level''}\label{figsmoothmatfactorone}} \hspace{.25in}
\parbox{2.5in}{\includegraphics[width=0.42
\textwidth]{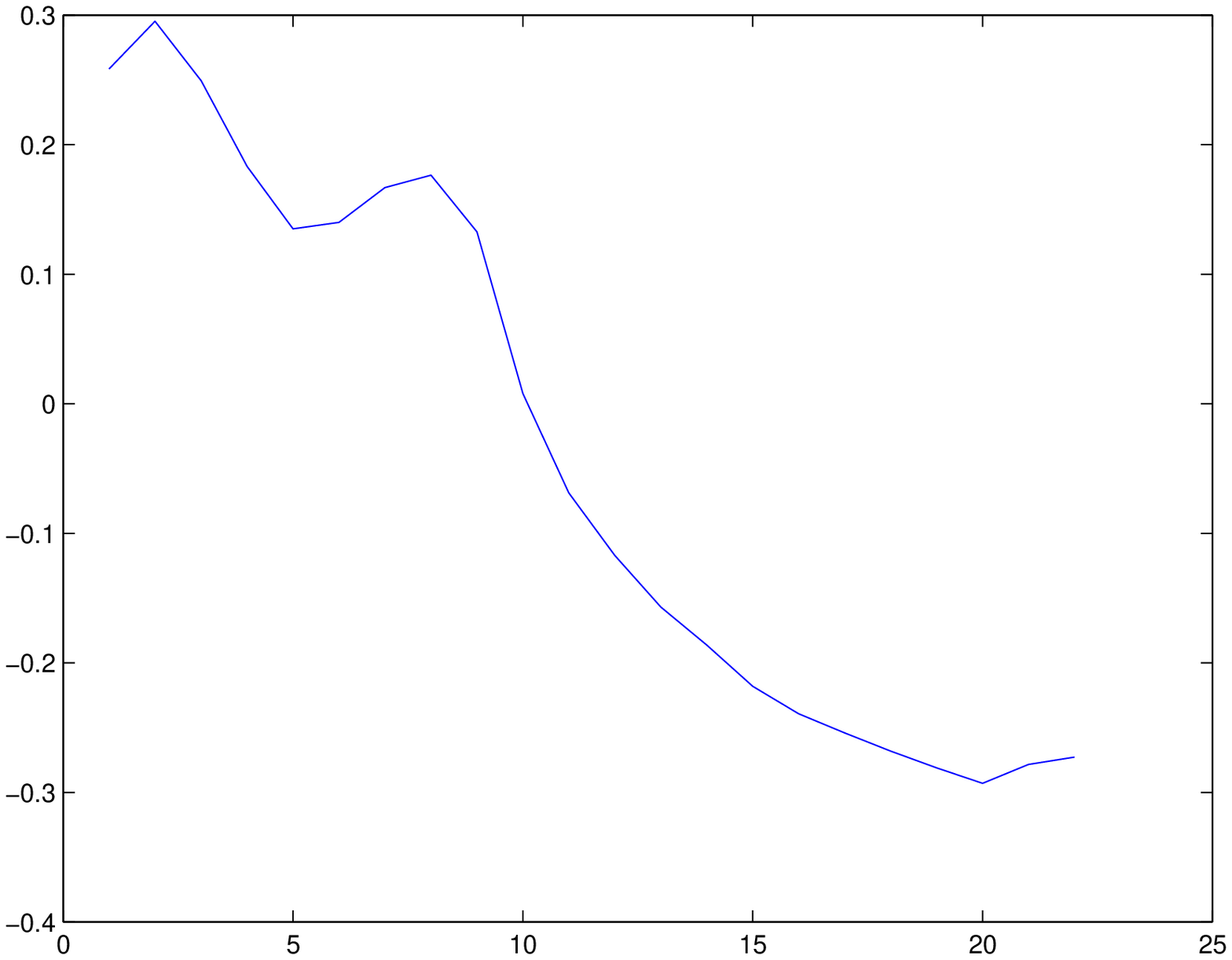} \caption{Second eigenvector
''spread''}\label{figsmoothmatfactortwo}}
\end{center}
\end{figure}

Using the same data set as above, we calibrate a covariance matrix under
smoothness constraints. The resulting matrix is plotted in figure (\ref
{figsmoothmatfactormat}).
In figure (\ref{figsmoothmatfactorone}) we plot the eigenvectors of this
matrix.
The first vector has a level shape while the second one is close
to a spread of rates. We can notice that this purely market
implied covariance factor structure closely matches the results
obtained using estimates from historical data.

\section{Conclusion}

The methods described in this work are organized around one central
objective: the design of a true ''black-box'' calibration and
risk-management tool for classic multifactor interest rate models. In
particular, the performance guarantee given by the numerical methods used
here makes it possible to design a calibration procedure that does not
require numerical baby-sitting. Furthermore, the possibility of stabilizing
the calibration result should induce significant savings in hedging
transaction costs by suppressing the possibility of purely numerical
calibration hedging and hence P\&L hikes.

In practice however, two important obstacles remain in the design
of a ''Swiss army knife'' interest rate model: smile modelling and
rank reduction. It is at this point not possible to globally
calibrate the model to both the smile and the covariance
structure, instead, one has to apply a two-step procedure to first
calibrate the correct smile structure and then recover the
covariance information using the methods detailed here. This makes
it impossible to jointly optimize the calibration result on the
smile and the covariance structure (for smoothness, stability,
etc...). The second problem is rank reduction: numerical methods
for American-style securities pricing are only efficient for
models with a small number of factors. Empirical evidence suggests
that market covariance matrixes should have a rapidly decreasing
eigenvalues and the matrix calibrated from market data on Caps and
Swaptions display that behavior, hence the rank reduction is
essentially a numerical backward compatibility problem and recent
advances in quantization methods (see \cite {ball00}) or American
Monte-Carlo (see \cite{long98} for example) make it reasonable to
believe that this limitation will eventually be lifted.

\bibliographystyle{agsm}
\bibliography{MainPerso}

\end{document}